\documentclass{article} 

\usepackage{arxiv}

\usepackage{amsmath,amsfonts,amsthm,amsfonts} 
\usepackage{breqn}
\usepackage[sort&compress,numbers,comma]{natbib}
\usepackage{hyperref}
\hypersetup{
	colorlinks = true,
	linkcolor = blue,
	anchorcolor = blue,
	citecolor = blue,
	filecolor = blue,
	urlcolor = blue,
}
\usepackage{esdiff}
\usepackage{graphicx,epstopdf,pgfplots,caption,subcaption}
\usepackage{appendix}
\usepackage[]{algorithm2e}

\usepackage{cleveref}
\Crefname{chapter}{Chapter}{Chapters}
\Crefname{section}{Section}{Sections}
\Crefname{figure}{Figure}{Figures}

\usepackage{autonum}

\renewcommand{\vec}[1]{\boldsymbol{#1}}
\newcommand{\p}[0]{\partial}

\newcommand{\bmth}[1]{\mbox{\boldmath $#1$}}
\newcommand{\grad}{\bmth{\nabla}}

\usepackage[labelfont=bf,font={tiny},margin=0.5cm]{caption}

\title{A review of one-phase Hele--Shaw flows and a level-set method for non-standard configurations}

\author{{\large Liam C.~Morrow$^{1,2}$, Timothy J.~Moroney$^2$, Michael C. Dallaston$^2$ and Scott W.~McCue$^2$}\\
	\vspace{0.cm} \\
	$^1$Department of Engineering Science, University of Oxford, Oxford OX1 3PJ, United Kingdom
	\vspace{0.10cm} \\
	$^2$School of Mathematical Sciences, Queensland University of Technology, Brisbane, QLD, 4001, Australia}

\begin{document}
	
\maketitle

\date{\today}

\begin{abstract}
The classical model for studying one-phase Hele--Shaw flows is based on a highly nonlinear moving boundary problem with the fluid velocity related to pressure gradients via a Darcy-type law.  In a standard configuration with the Hele--Shaw cell made up of two flat stationary plates, the pressure is harmonic.  Therefore, conformal mapping techniques and boundary integral methods can be readily applied to study the key interfacial dynamics, including the Saffman--Taylor instability and viscous fingering patterns.  As well as providing a brief review of these key issues, we present a flexible numerical scheme for studying both standard and non-standard Hele--Shaw flows. Our method consists of using a modified finite difference stencil in conjunction with the level set method to solve the governing equation for pressure on complicated domains and track the location of the moving boundary. Simulations show that our method is capable of reproducing the distinctive morphological features of the Saffman--Taylor instability on a uniform computational grid. By making straightforward adjustments, we show how our scheme can easily be adapted to solve for a wide variety of non-standard configurations, including cases where the gap between the plates is linearly tapered, the plates are separated in time, and the entire Hele--Shaw cell is rotated at a given angular velocity.

\vspace{1.5ex}
\end{abstract}

\section{Introduction}

Viscous fingering patterns that develop in a Hele--Shaw flow are very well studied in fluid dynamics. These patterns, which arise due to the Saffman--Taylor instability \citep{Saffman1958}, occur when a viscous fluid that fills a gap between two narrowly separated parallel plates is displaced by a less viscous fluid, which is injected into (or withdrawn from) the cell. Provided these two fluids are immiscible, an interface forms that is usually unstable and develops visually striking patterns characterised by their branching morphology. As the governing equation for the velocity of the viscous fluid is the same as Darcy's law, Hele--Shaw flow can be interpreted as a two-dimensional paradigm for flow through a homogeneous porous medium. Further, the Hele--Shaw framework has also been used to model interfacial instabilities appearing in other scenarios including bacterial colony growth \citep{Ben1990}, crystal solidification \citep{Li2004}, random walk simulations, \citep{Liang1986}, and the flow of electrolytes \citep{Mirzadeh2017,Gao2019b}.  We refer the reader to Refs~\citep{Casademunt2004,Homsy1987,Mccloud1995,Vasilev2009} for a historical summary and comprehensive overview of the broad applicability of the Hele--Shaw model.

If we assume that the viscosity of the less viscous fluid (air, say) can be neglected entirely, then the classical model for flow in the more viscous fluid, $\Omega(t)$, is the one-phase moving boundary problem
\begin{equation}
\boldsymbol{v} = -\frac{b^2}{12 \mu} \grad p,
\quad \nabla \cdot \boldsymbol{v}=0,
\quad \boldsymbol{x} \in \Omega(t),
\label{eq:standard1}
\end{equation}
where $\boldsymbol{v}$ is the fluid velocity (averaged across the distance $b$ between the parallel Hele--Shaw plates), $p$ is the fluid pressure and $\mu$ is the constant viscosity, together with the boundary conditions
\begin{equation}
p=-\gamma \kappa + \,\mbox{constant},
\quad
v_n = -\frac{b^2}{12 \mu} \frac{\partial p}{\partial n},
\quad
\boldsymbol{x} \in \partial \Omega(t),
\label{eq:standard2}
\end{equation}
where $\gamma$ is the surface tension, $\kappa$ is the signed curvature of $\partial\Omega$, defined to be positive if the interface is convex from the side of the more viscous fluid, and $v_n$ is the normal speed of the interface.  Typically the flow is driven by injection or withdrawal of fluid through a point or at infinity.  This original model for two immiscible fluids is described by Saffman \& Taylor \citep{Saffman1958} in 1958, except that for the most part they do not neglect the flow details of the less viscous fluid.

The one-phase Hele--Shaw model that we are concerned with has been applied to three main configurations, namely an expanding bubble of air into an infinite body of fluid, a contracting finite blob of fluid, and the displacement of viscous fluid in a Hele--Shaw channel.  In each of these three scenarios, the fluid boundary is unstable (the Saffman--Taylor instability), and a typical outcome involves portions of the interface propagating increasingly faster than other portions, in some cases leading to a striking fingering pattern at the boundary.  For the special zero-surface-tension case (also known as Laplacian growth), a host of mathematical studies based mostly on conformal mapping, conserved moments and the Baiocchi transform have highlighted the possible scenarios for this ill-posed model, including exact solutions and finite-time blow-up \citep{Crowdy2004,Cummings1999b,Cummings2004,Dallaston2012,Entov1995,Hohlov1993,Hohlov1994,Howison1986,Howison1986b,Howison1986c,Lacey1982,Mineev--Weinstein1998,Mineev--Weinstein2000}.
For the more physically realistic nonzero surface tension case (which is well-posed), the broader strategies to study this problem include stability analysis \citep{Miranda1998,Paterson1981}, small-surface-tension asymptotics \citep{Ceniceros2000b,tanveer_2000_surprises}, employing harmonic moments and conserved quantities \citep{Leshchiner2010} and fully numerical methods mostly with boundary integral methods \citep{Ceniceros1999,Dai1993,Hou1994,Hou2001,Kelly1997,Nie1998} but also the level set formulation \citep{Hou1997}.
While we shall devote much of our attention in this article to certain non-standard variations of (\ref{eq:standard1})-(\ref{eq:standard2}), we do not provide any commentary on how the boundary conditions (\ref{eq:standard2}) may be altered by considering additional physical effects on the boundary apart from surface tension, including the effects of a dynamic contact angle, thin wetting films and the related issue of kinetic undercooling \citep{Anjos2013,Anjos2015,Anjos2021,Chapman2003,Dallaston2013,Dallaston2014,ebert_2007,Park1984,Pleshchinskii2002,Saffman1986,Xie2019,Xie2021}.
Similarly, we do not review non-Newtonian flows, which themselves are well-studied \citep{aronsson_1992,Fast2001,Fontana2014,Kondic1998,McCue2011,Sader1994}.
Finally, our focus is on time-dependent problems and so we are not intending to review the extensive literature on travelling wave problems involving a steadily propagating finger \citep{Chapman1999,Combescot1986,Gardiner2015a,Gardiner2015b,Hong1986,Mclean1981,Saffman1986,Shraiman1986,Tanveer1987c,Vanden1983} or bubble \citep{Green2017,Hong1988a,Lustri2020,Tanveer1987a,Tanveer1987b,Tanveer1989,Vasconcelos1994,vasconcelos_2001}.

In recent years, there has been increased interest in studying how variations to the classic Hele--Shaw model influence the development of viscous fingering patterns. Many of these studies consider the effect of imposing a time-dependent injection rate, specifically to control or reduce the growth of fingers \citep{Arun2020,Beeson2017,Coutinho2020,Dias2010,Dias2012,Dias2010b,Gin2021,Li2009}. Further, much attention has been devoted to manipulating the geometry of the Hele--Shaw cell. One of the earliest examples of this approach is by \citet{Zhao1992}, who considered the classic Saffman--Taylor experiment \citep{Saffman1958} and linearly tapered the gap between the plates in the direction of the fluid flow. Since this experiment, numerous studies have been performed to generate further insight into how the taper angle influences viscous fingering \citep{Al2013,Al2012,Bongrand2018,Jackson2017,Lu2020,Morrow2019}. Other popular physical alterations to the Hele--Shaw cell include uniformly separating the plates in time \citep{Entov1995,Eslami2020,Lindner2005,Nase2011,Shelley1997,Vaquero2019,Zheng2015}, rotating the entire Hele--Shaw cell at a given angular velocity \citep{Anjos2017,Carrillo1996,Entov1995,Schwartz1989}, or replacing one of the plates with an elastic membrane \citep{Al2013b,Cuttle2020,Fontana2021,Lister2013,McCue2018,Pihler2012,Pihler2013,Pihler2014,Pihler2018}. All of these configurations have been shown to produce patterns distinct from traditional Saffman--Taylor fingering.

One of the most commonly used analytical tools for studying both standard and non-standard Hele--Shaw flow is linear stability analysis. For the standard configuration, \citet{Paterson1981} showed that modes of perturbation to the circular solution become successively unstable as the bubble expands, predicting the most unstable wave number for a given bubble radius. Further, linear stability analysis has also been used to derive injection rates to control \citep{Li2007} or minimise \citep{Dias2012} the development of viscous fingering. For non-standard Hele--Shaw flow, linear stability analysis provides insight into how manipulating the geometry of the cell influences the development of viscous fingers, including when the plates are tapered \citep{Al2013,Al2012}, rotating \citep{Carrillo1996}, or are being separated \citep{Shelley1997}. While linear stability analysis is a flexible tool that leads to analytic predictions \citep{Dias2010,Gin2015a,Li2009}, it only leads to an accurate description of solutions for small time. As such, this restriction increases the need for flexible and accurate numerical methods that can be used to understand the full nonlinear behaviour of these problems.

Computing numerical solutions to Hele--Shaw flow (and related moving boundary problems) can be a challenging task, as interfacial patterns develop, which requires solving the governing equations in complicated moving domains. Such approaches can be classified as either front tracking, where the interface is solved for explicitly, or front capturing, where the interface is represented implicitly. For the classic Hele--Shaw problem, as the pressure of the viscous fluid satisfies Laplace's equation, the most popular choice is the boundary integral method (also known as the boundary element method), which is classified as a front tracking method. In particular, the boundary integral method has been used to solve the classic one-phase Hele--Shaw problem \citep{Dai1993,Li2007,Li2009}, as well as two-phase flow \citep{Jackson2015,Power2013}, problems for which the plates are uniformly separated in a time-dependent fashion \citep{Shelley1997,Zhao2018,Zhao2021}, and Hele--Shaw flow in channel geometry \citep{Degregoria1986}. However, for non-standard Hele--Shaw configurations, the pressure may no longer be harmonic and the boundary integral method becomes a less suitable tool.  Another disadvantage of front tracking methods is that the mesh may need to be regenerated as the interface evolves, in which case care must be taken to avoid mesh distortion effects.

A popular alternative to the boundary integral method is the level set method, which represents the interface implicitly as the zero level set of a higher dimensional hypersurface \citep{Osher1988}. A commonly cited advantage of the level set method is that it can easily handle complicated interfacial behaviour such as the merging and splitting of interfaces. Another, more pertinent, advantage of the level set method is that it can describe the formation of complicated interfacial patterns (such as those that occur in Hele--Shaw flow) on a uniform grid, eliminating the need to re-mesh as the interface evolves. One of the most significant drawbacks of the level set method is that it can suffer from mass loss/gain in regions where the mesh is under resolved. However, this issue can be mitigated by using the particle level set method \citep{Enright2002}, which uses massless marker particles to correct the location of the interface when mass is lost/gained. The level set method is a popular tool for studying moving boundary problems in fluid dynamics, and has been used to investigate interfacial instabilities that occur in Stefan problems \citep{Chen1997,Gibou2003} and Hele--Shaw flow \citep{Hou1997,Lins2017}.  Also, we have applied this method to these applications, in particular to conduction-limited melting of crystal dendrites \citep{Morrow2019b}, bubbles shrinking and breaking up in a porous medium \citep{Morrow2019c}, and bubbles expanding in various Hele-Shaw configurations \citep{Morrow2019}.  We refer to Refs~\citep{Gibou2018,Osher2001,Sethian2003} for more information about the level set method, including details regarding implementation and applications.

While the level set method is used to implicitly represent the location of the interface, to numerically simulate Hele--Shaw flow we are also required to determine the pressure within the viscous fluid, which involves solving a partial differential equation in a complicated domain that changes in time. When applying the boundary integral method for the classic Hele--Shaw problem, the solution to Laplace's equation can be expressed in terms of Green's functions. As such, the problem is reformulated as an integro-differential equation, and nodes need only be placed on the fluid-fluid interface. An alternative choice is to solve for the pressure using the finite difference method, which can be modified to solve problems on complicated domains when coupled with level set functions that describe the location of the interface \citep{Coco2013,Gibou2002}. An advantage of this approach is that the finite difference method can be easily adapted to wide classes of partial differential equations. Further, while the boundary integral method can easily handle non-trivial far-field boundary conditions, their inclusion into the finite difference stencil is not so straightforward. One solution to overcome this difficulty is to use a very large computational domain, but this in turn results in significantly longer computational times. Another, more elegant, solution is to use a Dirichlet-to-Neumann map \citep{Givoli2013}, which has been shown to accurately capture the far-field boundary condition even when the interface is relatively close to the curve on which the Dirichlet-to-Neumann map is applied \citep{Morrow2019c,Morrow2019b,Morrow2019}.

In this work, we provide a brief review of the one-phase Hele--Shaw model, touching on the use of complex variable and conformal mapping techniques as well as the mathematical consequences of including or excluding surface tension in the model.  We focus on the three well-studied scenarios, namely an expanding bubble, a contracting blob and displacement of fluid in a linear channel. Our approach is to write down a generalised model that allows for a number variations of the standard approach, including a time-dependent flow rate, a spatially and/or time-dependent gap between the plate, or rotating plates.  We then present a flexible numerical framework for solving this generalised model \citep{Morrow2019c,Morrow2019b,Morrow2019}. Our scheme is based on the work of \citet{Chen1987} and \citet{Hou1997}, and uses a level set based approach to track the location of the liquid-air interface. There are several novel aspects of our numerical framework. The first is that our scheme overcomes the limitations of the boundary integral method in that it can easily solve Hele--Shaw flow in non-homogeneous media, i.e.~where the plates are not parallel. Second, by representing the interface implicitly by a higher dimensional level set function, we are able represent the complicated interfacial patterns easily on a uniform mesh. By performing a series of simulations, we show that our numerical solutions are able to reproduce the morphological features of viscous fingering in a Hele--Shaw cell.  Further, by making straightforward adjustments, we show that our scheme can easily be modified for a wide range of non-standard Hele--Shaw configurations, including where the plates are linearly tapered, uniformly separated in time, or rotated. For all the configurations considered, our numerical solutions are shown to compare well with previous simulations and experiments.

\section{Review of one-phase Hele--Shaw model} \label{sec:HeleShawModel}

\subsection{Generalised Hele--Shaw model}

We consider a generalised one-phase model of Hele--Shaw flow where the gap between the plates is either spatially or temporally dependent such that $b \to b(\boldsymbol{x}, t)$ and the Hele--Shaw plates can rotate with angular velocity $\bar{\omega}$.  We suppose an inviscid bubble is injected into the viscous fluid at rate $Q(t)$, and denote the domain occupied by the inviscid fluid to be $\Omega(t)$.  The interface separating the inviscid bubble and the viscous fluid is denoted by $\partial \Omega(t)$.  Is is commonplace with Hele-Shaw flows, we consider a depth averaged model that comes about from averaging Stokes flow over the gap between the plates, which itself is assumed to be small.

Denoting $P$, $\mu$ and $\rho$ as the pressure, viscosity and density of the viscous fluid, respectively, and denoting the angular velocity of the Hele--Shaw cell by $\bar{\omega}$, the governing equations for the velocity of the viscous fluid are
\begin{align}
\boldsymbol{v} &= -\frac{b^2}{12 \mu} (\grad P - \bar{\omega}^2 \rho r \boldsymbol{e}_r) ,  &\boldsymbol{x} &\in \mathbb{R}^2 \backslash \Omega(t),  \label{eq:Governing1} \\
\grad \cdot (b \boldsymbol{v}) &= - \frac{\partial b}{\partial t},  &\boldsymbol{x} &\in \mathbb{R}^2 \backslash \Omega(t),  \label{eq:Governing2}
\end{align}
where $r = |\vec{x}|$.  Equation \eqref{eq:Governing1} is Darcy's law modified to include the rotational effects of the Hele--Shaw cell, while \eqref{eq:Governing2} ensures that the mass of the fluid is conserved. Defining a reduced pressure according to $p = P - \bar{\omega} \rho r^2 / 2$ and then substituting \eqref{eq:Governing1} into \eqref{eq:Governing2} generates the governing equation for pressure,
\begin{align}
\grad \cdot \left(  \frac{b^3}{12 \mu} \grad p \right)  &= \frac{\partial b}{\partial t}, &&\boldsymbol{x} \in \mathbb{R}^2 \backslash \Omega(t)       \label{eq:HeleShaw10}.
\end{align}
When the gap between the plates is both spatially and temporally uniform, \eqref{eq:HeleShaw10} reduces to Laplace's equation $\nabla^2 p = 0$. We have two boundary conditions on the fluid-fluid interface given by
\begin{align}
p &= -\gamma \left(  \kappa + \frac{2}{b} \right) - \frac{\rho \bar{\omega}^2 r^2}{2},  &\boldsymbol{x} &\in \partial \Omega(t)       \label{eq:HeleShaw20}, \\
v_n &= -\frac{b^2}{12 \mu} \frac{\partial p}{\partial n},       &\boldsymbol{x} &\in \partial \Omega(t)     		 			\label{eq:HeleShaw30},
\end{align}
where $\gamma$ is the surface tension, $\kappa$ is the signed curvature of $\partial \Omega$, $\rho$ is the density of the viscous fluid, and $v_n$ is the normal speed of the interface. The dynamic boundary condition \eqref{eq:HeleShaw20} incorporates both the effects of surface tension as well as the rotation of the Hele--Shaw plates. The kinematic boundary condition \eqref{eq:HeleShaw30} relates the velocity of the viscous fluid with the normal velocity of the interface. We also have the far-field boundary condition
\begin{align}
\frac{b^3}{12 \mu}\frac{\partial p}{\partial r} &\sim -\frac{Q}{2 \pi r} + \frac{1}{2} r \frac{\partial b}{\partial t}, && r \to \infty,					\label{eq:HeleShaw40}
\end{align}
which acts as a source/sink term at infinity. For $Q > 0$ ($Q < 0$), this condition corresponds to the bubble area expanding (contracting) at rate $Q$. The inclusion of the $\p b / \p t$ in \eqref{eq:HeleShaw40} comes about from the non-homogenous term in \eqref{eq:HeleShaw10}, and ensures the change of rate of volume of the bubble is $Q$.

To nondimensionalise \eqref{eq:HeleShaw10}-\eqref{eq:HeleShaw40}, we introduce $r_0$ as the average initial radius of the bubble and $Q_0$ as the average injection rate over the duration of a simulation, and $b_0 = b(0,0)$.  Then space, time, pressure, and velocity are scaled according to 
\begin{align}
\hat{\vec{x}}=\frac{\vec{x}}{r_0}, \quad \hat{t}=\frac{t}{T}, \quad \hat{b}=\frac{b}{b_0},
\quad \hat{p}=\frac{b_0^2T}{12\mu r_0^2}p, \quad \hat{\vec{v}}=\frac{T}{r_0}\vec{v},
\label{eq:scaledvariables}
\end{align}
where $T$ is a representative time-scale.  Dropping the hats and retaining our original variable names, \eqref{eq:HeleShaw10}-\eqref{eq:HeleShaw40} become
\begin{alignat}{3}
\nabla \cdot \left( b^3 \nabla p \right)  &= \frac{\partial b}{\partial t}, &\boldsymbol{x} &\in \mathbb{R}^2 \backslash \Omega(t), \label{eq:HeleShaw1}\\
p &= -\sigma \left(  \kappa + \frac{2 R_0}{b} \right) - \omega^2 r^2, \qquad \qquad  &\boldsymbol{x} &\in \partial \Omega(t), \label{eq:HeleShaw2} \\
v_n &= -b^2 \frac{\partial p}{\partial n},       &\boldsymbol{x} &\in \partial \Omega(t), \label{eq:HeleShaw3}\\
b^3 \frac{\partial p}{\partial r} &\sim -\frac{Q}{2 \pi r} + \frac{1}{2} r \frac{\partial b}{\partial t} & r &\to \infty, \label{eq:HeleShaw4}
\end{alignat}
where $\sigma = b_0^2 T\gamma / 12 \mu r_0^3$, $R_0 = r_0/b_0$, and $\omega^2 = \rho b_0^2 T\bar{\omega}^2  / 24 \mu$.  For this configuration, an appropriate time-scale could be $T=r_0^2b_0/Q_0$, in which case the dimensionless average injection rate would become $Q_0=1$.

In addition to this expanding bubble problem, we shall also be concerned with two other scenarios, namely the blob geometry, where viscous fluid occupies $\Omega(t)$ and is withdrawn from a point or the cell rotates around a perpendicular axis, and the channel geometry, where viscous fluid occupies a long rectangular channel and is displaced by the inviscid fluid that is injected at one end.  For these two scenarios, modifications to (\ref{eq:HeleShaw1})-(\ref{eq:HeleShaw4}) will be made as appropriate.

\subsection{Complex variable formulation}\label{sec:complex}

Before outlining our numerical scheme in Section~\ref{sec:NumericalScheme}, we take some time to illustrate some of the mathematical properties of the Hele-Shaw problem, especially in the special case of zero surface tension.  This mathematical exploration, which relies heavily on complex variable theory and conformal mapping, is based on many previous studies in this spirit
\citep{Crowdy2004,Cummings1999b,Cummings2004,Dallaston2012,Hohlov1993,Hohlov1994,Howison1986,Howison1986b,Howison1986c,Mineev--Weinstein1998,Mineev--Weinstein2000}.
To keep the discussion contained and to connect with numerical simulations described later in this paper, we restrict ourselves to examples of three geometries (the expanding bubble problem, the blob problem and the channel geometry).

In the standard set-up in which $b=1$ (parallel, stationary plates) and $\omega=0$ (no rotation), equations \eqref{eq:HeleShaw1}-\eqref{eq:HeleShaw4} reduce to
\begin{alignat}{3}
\nabla^2 p &= 0, &\boldsymbol{x} &\in \mathbb{R}^2 \backslash \Omega(t), \label{eq:HeleShawstandard1}\\
p &= -\sigma\kappa, \qquad \qquad  &\boldsymbol{x} &\in \partial \Omega(t), \label{eq:HeleShawstandard2} \\
v_n &= - \frac{\partial p}{\partial n},       &\boldsymbol{x} &\in \partial \Omega(t), \label{eq:HeleShawstandard3}\\
\frac{\partial p}{\partial r} &\sim -\frac{Q}{2 \pi r}, & r &\to \infty. \label{eq:HeleShawstandard4}
\end{alignat}
It is instructive to reformulate this problem using complex variable methods as follows.  Given the fluid pressure $p$ satisfies Laplace's equation \eqref{eq:HeleShawstandard1}, it can be interpreted as the negative real part of an analytic function $W(z,t)=-p(x,y,t)+\mathrm{i}\psi(x,y,t)$ of the complex variable $z=x+\mathrm{i}y$.  Here, $W$ is acting as a complex potential, while $\psi$ is a streamfunction.

\begin{figure}
\centering
\includegraphics{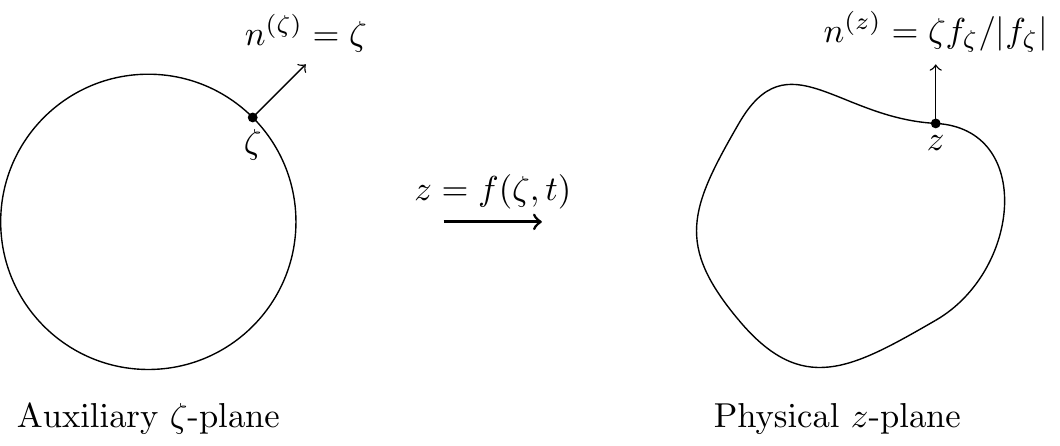}
\caption{A schematic diagram indicating the time-dependent complex mapping $f(\zeta,t)$ from the auxiliary $\zeta$ plane to the physical $z (=x+\mathrm i y)$ plane.  The interface $\delta\Omega(t)$ is the image of the unit circle $|\zeta|=1$, and the complex representation of the unit normal vector can be expressed in terms of the derivative of $f$.}
\label{fig:complex_schematic}
\end{figure}

Further, there exists a time-dependent conformal map $z=f(\zeta,t)$ from the unit disc in the plane of an auxiliary variable $\zeta$ to the fluid region in the $z$-plane (i.e., $\mathbb{R}^2 \backslash \Omega(t)$) and the unit circle $|\zeta|=1$ to the fluid interface $\partial\Omega(t)$, as depicted schematically in \Cref{fig:complex_schematic}.  The map will be univalent (that is, one-to-one) and analytic in the unit disc except for at a single point, which we choose to be $\zeta=0$, that represents $z\to\infty$.  In the limit $\zeta\rightarrow 0$, we have $f\sim a(t)/\zeta$.  By fixing a rotational degree of freedom we force $a(t)$ to be real.  Now the complex potential $W(z,t)$ is also an analytic function of $\zeta$ and so we write $w(\zeta,t)=W(f(\zeta,t),t)$.  Given the far-field condition \eqref{eq:HeleShawstandard4}, which implies $W\sim (Q/2\pi)\log z$ as $|z|\rightarrow\infty$, we now have the local behaviour $w\sim -(Q/2\pi)\log\zeta$ as $\zeta\rightarrow 0$.

To formulate the kinematic condition (\ref{eq:HeleShawstandard3}) in terms of the map $f$, it is useful to introduce some complex variable equivalents of standard concepts from vector algebra.  Firstly, a complex number can be used to represent a vector (with components given by the real and imaginary parts), such as the normal to an interface.  The unit normal to the unit circle $|\zeta|=1$ is $n^{(\zeta)} = \zeta$, and given the interface $\partial\Omega$ is the image of the unit circle under $z = f(\zeta,t)$, the normal $n^{(z)}$ in the $z$-plane is found by rotating $n^{(\zeta)}$ by the argument of $f_\zeta$, thus $n^{(z)} = \zeta f_\zeta/|\zeta f_\zeta|$ (see \Cref{fig:complex_schematic}).  Secondly, the equivalent of the dot product between two complex numbers $a$ and $b$ is $\Re\{a\overline b\}$.  The time derivative of the map $f_t$ at a point on the unit disc gives a velocity vector of a point on the interface $\partial\Omega$. Therefore the normal velocity $v_n$ of the interface $\delta\Omega$ as a function of $\zeta$ is given by $\Re\{f_t\overline{\zeta f_\zeta}\}/|\zeta f_\zeta|$, while the normal derivative $\partial p/\partial n$ is given by $-\Re\{\zeta W_z f_\zeta\}/|f_\zeta| = \Re\{\zeta w_\zeta\}/|f_\zeta|$.  This calculation allows the kinematic condition to be represented as
$$
\Re\{f_t\overline{\zeta f_\zeta}\}=\Re\{\zeta w_\zeta\},
\qquad |\zeta|=1.
$$
Now to reformulate the dynamic condition \eqref{eq:HeleShawstandard2} we note the curvature on the fluid boundary can be written as $\Re\{\zeta(\zeta f_\zeta)_\zeta\overline{\zeta f_\zeta}\}/|\zeta f_\zeta|^3$  on the unit circle.  Given \eqref{eq:HeleShawstandard2} and the logarithmic behaviour of $w$ as $\zeta\rightarrow 0$, we can write $w=-(Q/2\pi)\log\zeta-\sigma\mathcal{K}(\zeta,t)$, where $\mathcal{K}(\zeta,t)$ is an analytic function of $\zeta$ in the unit disc whose real part on the unit circle $|\zeta|=1$ is given by the curvature $\kappa$.  Combining these ideas, we arrive at the single governing equation
\begin{equation}
\Re\{f_t\overline{\zeta f_\zeta}\}=-\frac{Q}{2\pi}-\sigma\Re\{\zeta \mathcal{K}_\zeta\}
\qquad |\zeta|=1.
\label{eq:PGeqn}
\end{equation}
This equation is often referred to as the Polubarinova--Galin equation, especially when surface tension is ignored \cite{Gustafsson2006}.

\begin{figure}
	\centering
	\includegraphics[width=0.57\linewidth]{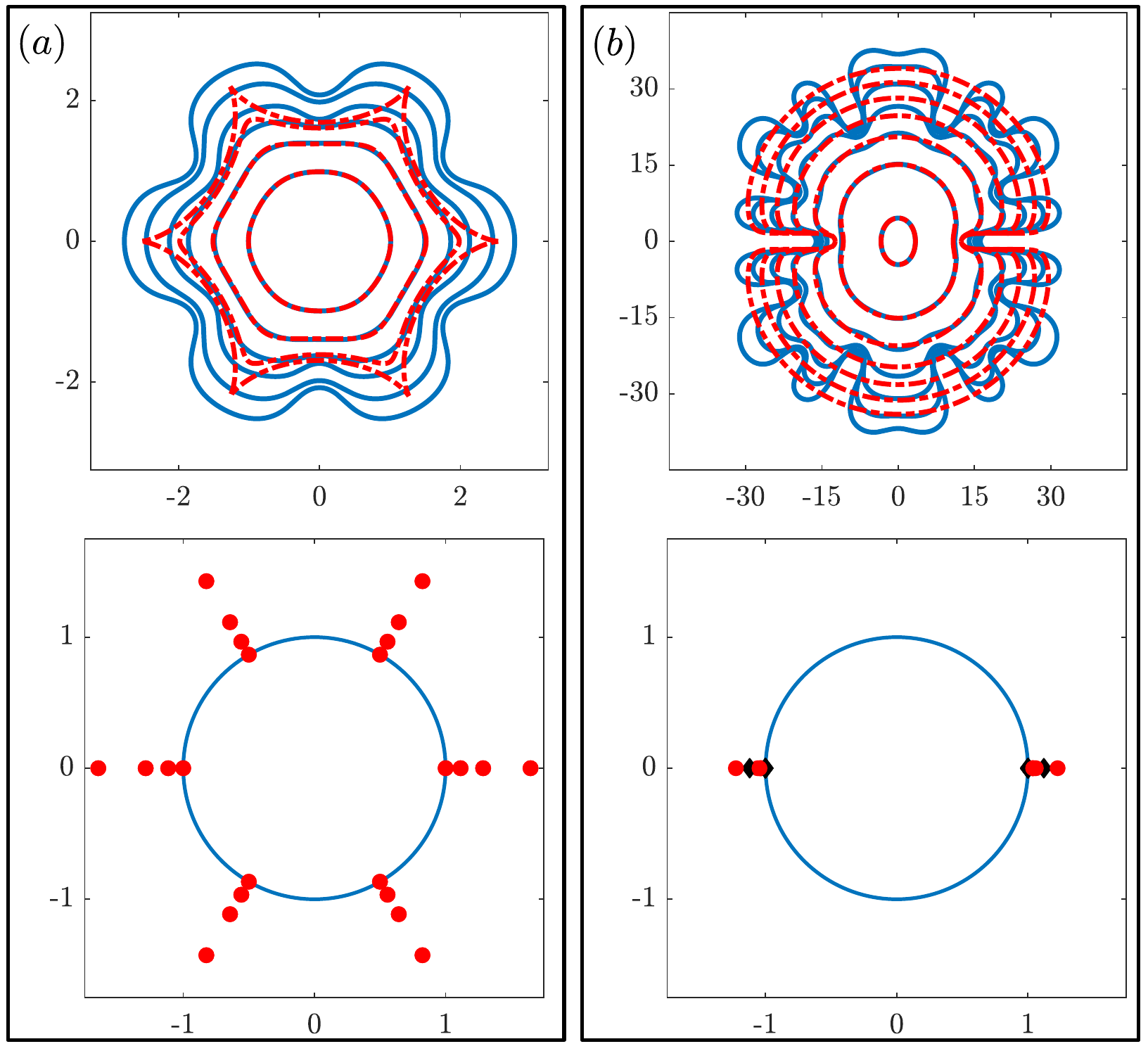} \\
	\includegraphics[width=0.57\linewidth]{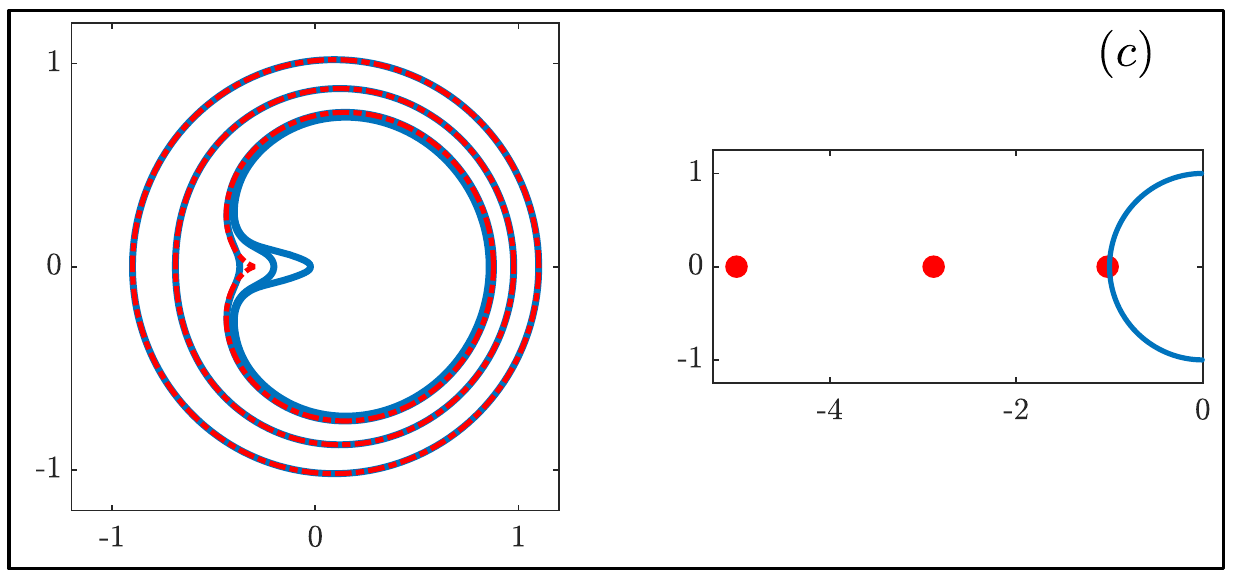}\\
	\includegraphics[width=0.57\linewidth]{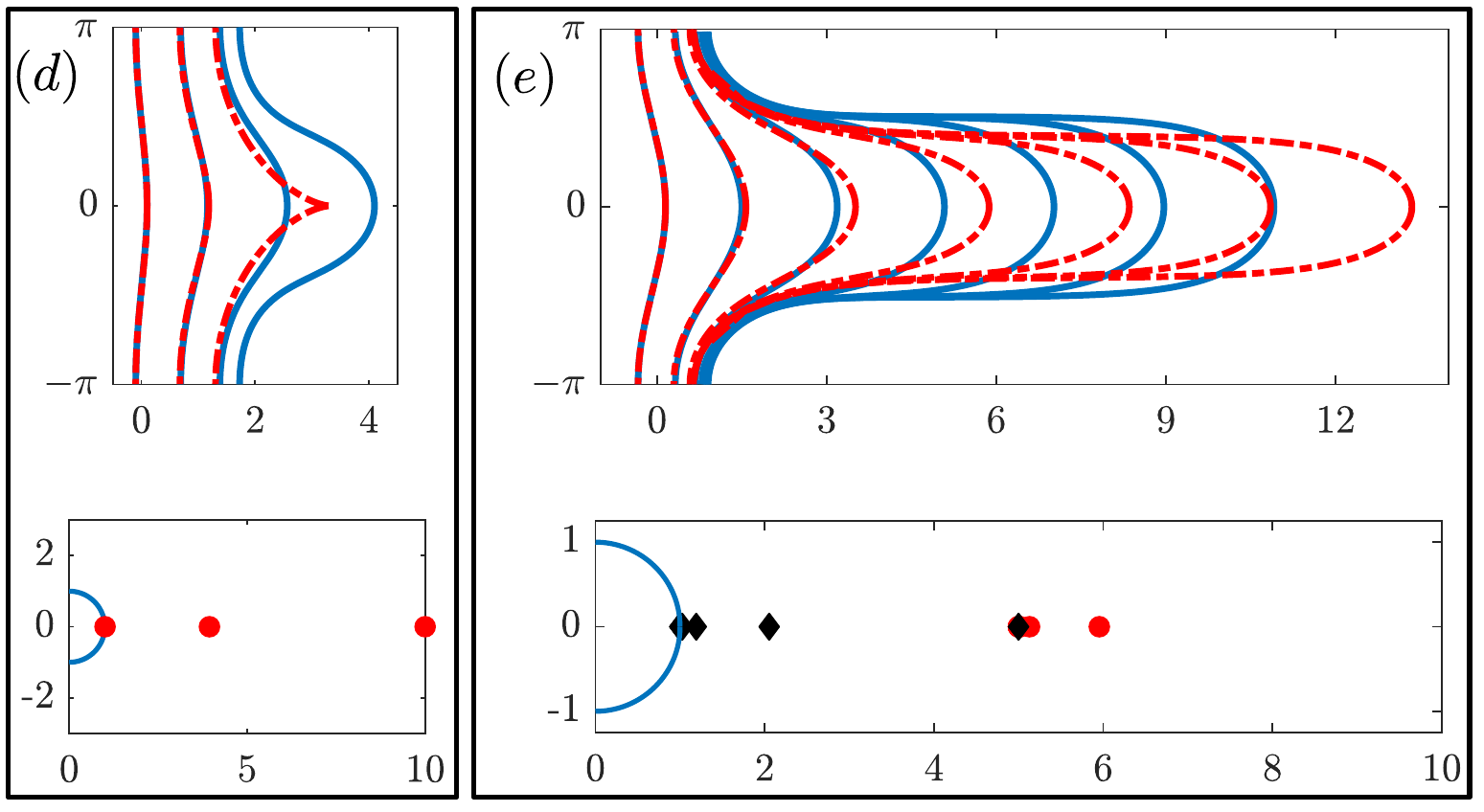}
	\caption{Solutions with and without surface tension.  The red dashed curves in (a)-(e) are zero-surface-tension solutions described by the five examples in \Cref{sec:complex}.  The solid blue curves are numerical solutions, including surface tension, for the same initial conditions.  For the examples in (a), (c) and (d), the zero-surface-tension solutions involve a form of finite-time blow-up characterised by cusps forming on the interface; the inclusion of surface tension regularises these singularities, allowing the full solution to continue past these blow-up times.  For the examples in (b) and (e), blow-up in the zero-surface-tension solution is prevented by the presence of a logarithmic singularity; here the numerical solution with small surface tension remains close to the zero-surface-tension solution for small time and then deviates away so that the long-time behaviour is different.  Each case includes a sketch of the $\zeta$-plane with the unit circle and critical points and logarithmic singularities indicated by solid red dots and black diamonds, respectively.  Note in (b) we do not plot the critical points at $t=0$ as they are outside the field of view here. For $(a)$-$(e)$, as time increases, the critical points and logarithmic singularities move towards the unit circle.}
	\label{fig:ZST}
\end{figure}

We shall briefly outline five illustrative examples, chosen to demonstrate the key features for the special case in which surface tension is ignored.  We shall later refer back to these examples when we implement our numerical scheme with surface tension included.  First, we shall consider the mapping $f=a(t)/\zeta+b(t)\zeta^N$, $N\geq 2$ \cite{Howison1986}.  By substituting into (\ref{eq:PGeqn}) with $\sigma=0$, we find that $a$ and $b$ must satisfy the coupled system of ODEs $a\dot{a}-Nb\dot{b}=Q/2\pi$, $N\dot{a}b-a\dot{b}=0$.  Say, for definiteness, $a(0)=1$ and $b(0)=\epsilon$, then the second of these equations gives $b=\epsilon a^N$, while the first equation integrates to give the time-dependence
$$
t=\frac{\pi}{Q}\left(a^2-1-\epsilon^2N(a^{2N}-1)\right).
$$
Thus we have an exact solution, as shown by the red dashed curves in the top panel of \Cref{fig:ZST}(a).   The innermost curve is the initial bubble boundary.  Here we have chosen $N=5$, $Q=2\pi$ and $\epsilon=0.01$, so this inner curve is a circle with a six-fold perturbation.  As time increases, the bubble expands and starts to develop six small fingers.

It is of interest to track the critical points, $\zeta=\zeta^*$, which are the points at which $f_\zeta=0$.  For this example, $\zeta^*=(1/\epsilon N a^{N-1})^{1/(N+1)}$.  Clearly there are
$N+1$ critical points that are equally spaced along a circle that is outside the unit circle in the $\zeta$-plane.  As time evolves, each of these critical points moves in a straight line towards the origin and intersects the unit circle when $|\zeta^*|=1$, ie $a=1/(\epsilon N)^{1/(N-1)}$.  We can compute the exact time that this occurs, namely
$$
t^*=\frac{\pi}{Q}\left(\frac{N-1}{N(\epsilon N)^{2/(N-1)}}-1+\epsilon^2N
\right).
$$
For the case in \Cref{fig:ZST}(a), for which $N=5$, we can see the six critical points (red dots) in the bottom panel moving towards the unit circle.  At $t=t^*$, there is finite-time blow-up, characterised by six cusps of order $3/2$ along the bubble boundary, which we can see in the top panel of \Cref{fig:ZST}(a) (a cusp of order $3/2$ is characterised by a curvature singularity that appears locally like two branches of $y^2=x^3$ meeting at a cusp, suitably scaled and rotated).  The solution cannot be continued past $t=t^*$ as the conformal mapping ceases to be univalent.

The second zero-surface-tension example we consider is for $f=a(t)/\zeta+\log(\zeta-d(t))-\log(\zeta+d(t))$ \cite{Howison1986}.  Again, the geometry is an expanding bubble, however this time the behaviour is qualitatively different.
By substituting this map for $f$ into (\ref{eq:PGeqn}) with $\sigma=0$, we find that $a$ and $d$ satisfy the coupled system:
$$
\dot{a}=\frac{Q}{2\pi}\frac{ad^4-a+4d}{a^2d^4-(2d-a)^2},
\quad
\dot{d}=-\frac{Q}{2\pi}\frac{d(d^4-1)}{a^2d^4-(2d-a)^2}.
$$
For initial conditions, we choose both $a(0)>d(0)>1$ so the denominators are initially positive, which means that $\dot{a}>0$ and $\dot{d}<0$ for small time.  To determine the precise time-dependent behaviour, one option is to integrate this system numerically, which demonstrates that $a(t)$ continues to increase while $d(t)$ decreases towards $d=1$ as $t$ increases.  To make progress analytically, by dividing one equation by the other, we can also derive a first-order ode with exact solution
$$
a=\frac{1}{d}\left[\ln\left(\frac{(d(0)^2-1)(d^2+1)}{(d(0)^2+1)(d^2-1)}\right)+a(0)d(0)\right],
$$
which shows that $a\sim -\ln(d-1)$ as $d\rightarrow 1^+$.  Clearly the map has logarithmic singularities at $\zeta_s=\pm d$ and critical points where $\zeta^*=\pm d\sqrt{a}/\sqrt{a-2d}$.  For sufficiently large time, we have $a-2d>0$ and so both $\zeta_s$ and $\zeta^*$ lie on the real $\zeta$-axis and move towards the unit circle as $t\rightarrow\infty$.  Since $|\zeta^*|\sim d+d^2/a$ as $a\rightarrow\infty$, we see the critical points $\zeta^*$ are further away from the unit circle than the logarithmic singularities $\zeta_s$, and therefore finite-time blow-up is avoided (each logarithmic singularity asymptotes to the unit circle, but does not cross it; see \citet{Howison1985}, and the discussion on the channel problem at the end of this section).

This second example is illustrated in \Cref{fig:ZST}(b) using $a(0)=4$, $d(0)=10/3$, $Q=2\pi$.  In the top panel, the exact solution is denoted by the dashed red curves.  Initially the bubble boundary looks oval in shape on this scale.  As time increases the interface expands, leaving two fjords behind, centred both the positive and negative $x$-axes.  In the bottom panel, both the critical points (red dots) and logarithmic singularities (black dots) are indicated in the $\zeta$-plane.  As just described, even though the critical points move towards the unit circle, they are further away than the logarithmic singularities; therefore, the logarithmic singularities have the effect of shielding the critical points and preventing blow-up.  An interesting observation is that each of the two fjords appears to take the shape of a classical Saffman-Taylor finger \citep{Saffman1958}.  To see this, note that on the unit circle near $\zeta=1$ we can derive a local analysis by setting $\zeta=1+\mathrm{i}\eta$, so that $f\sim a+\log(1-d+\mathrm{i}\eta)$.  Taking real and imaginary parts and then eliminating $\eta$ gives $x=\mathrm{constant}-\ln(\cos y)$, which is the famous Saffman-Taylor finger shape with width $\pi$.

The third example is probably the most well-known example of an exact solution in Hele--Shaw flow.  Here suppose the geometry is such that there is a blob of fluid in the Hele--Shaw cell, occupying a region $\Omega(t)$, surrounded by inviscid fluid.  If the fluid is withdrawn from a point in space (the origin, say), then the blob boundary contracts and we have the less viscous fluid displacing the more viscous fluid.  The governing equations (\ref{eq:HeleShawstandard1})-(\ref{eq:HeleShawstandard3}) apply in $\Omega(t)$, while (\ref{eq:HeleShawstandard1}) is replaced by $\partial p/\partial r\sim Q/2\pi r$ as $r\rightarrow 0$.  The map we have in mind here for this example is the quadratic map $f=a(t)\zeta+b(t)\zeta^2$, where the initial conditions $a(0)=1$, $b(0)=\epsilon\ll 1$ correspond to an initial blob boundary that is a perturbed circle \cite{Polubarinova1945}.  Substituting this map into the Polubarinova--Galin equation with $\sigma=0$ leads to a coupled system of integrable odes with the exact solution
$$
a^2b=\epsilon,\quad t=\frac{\pi}{Q}\left(1-a^2+2(\epsilon^2-b^2)\right).
$$
There is one critical point $\zeta^*=-a^3/2\epsilon$, which is initially located in the $\zeta$-plane at $-1/2\epsilon$ and moves towards the origin as time evolves and $a$ decreases.  At $t=t^*$ this critical point hits the unit circle, where
$$
t^*=\frac{\pi}{Q}\left(1+2\epsilon^2-\frac{9}{8}(2\epsilon)^{2/3}
\right),
$$
causing a cusp of order $3/2$ to form on the blob boundary.  This behaviour is illustrated in \Cref{fig:ZST}(c) for $\epsilon=0.1$ and $Q=2\pi$.  On the left panel, the blob boundary is represented by the dashed red curves.  Initially, this curve appears circular, as the perturbation is very small.  For intermediate times, the left portion of the boundary begins contracting faster than the remainder of the boundary until the cusp forms, corresponding to finite-time blow-up.  On the right panel of \Cref{fig:ZST}(c) the location of the fixed point in the $\zeta$-plane is represented (red dots) for the same three times that the boundary is drawn for in the left panel.  Here we see that the critical point touches the unit circle at the precise time that finite-time blow-up occurs.  Note that for polynomial maps like the one in this example, it has been proven that a cusp will always form before the interface reaches the sink \cite{Hohlov1993}; other explicit solutions exist whose boundary evolves to the location of the sink before or at the same time as cusp formation (see \cite{Richardson1972}, for example).  The time reversibility of the system (\ref{eq:HeleShawstandard1})-(\ref{eq:HeleShawstandard4}) in the absence of surface tension ($\sigma=0$) implies that the only initial condition that will lead to the removal of all fluid is a disc centred on the sink, that is, $f(\zeta,t) = a(t)\zeta$.

For completeness we include two more examples, providing only the key details.  These examples are for the geometry of flow in a Hele--Shaw channel, which we fix to be $2\pi$ units wide.  For the fourth example, the map takes for the $f=-\log\zeta+a(t)+b(t)\zeta$, with initial conditions $a(0)=0$, $b(0)=\epsilon\ll 1$, corresponding to a slightly perturbed flat interface \cite{Howison1985}.  This case is analogous to the first and third examples above.  The functions $a$ and $b$ satisfy the coupled system of odes with an exact solution
$$
a-\ln b=-\ln\epsilon, \quad t=\frac{\pi}{Q}\left(2a-b^2+\epsilon^2\right).
$$
There are critical points at $\zeta^*=1/b$ that move towards the origin as $b$ increases and ultimately intersect the unit circle at $t^*=\pi(2\ln(1/\epsilon)-1+\epsilon^2)/Q$, at which time a $3/2$ cusp forms on the interface.  This example is illustrated in \Cref{fig:ZST}(d), where the interface profiles are shown in the top panel as red dashed curves.  In the bottom panel the critical points are indicated (red dots).  Here $\epsilon=0.1$ and so the critical point is initially at $\zeta^*=10$ and ultimately hits the unit circle at $t=t^*$.

Finally, the fifth example is for $f=-\log\zeta+a(t)+\alpha\log(\zeta+d(t))$ with initial conditions $a(0)=0$, $d(0)=1/\epsilon\gg 1$ \cite{Howison1985}, which is analogous to the second example above.  There is a critical point at $\zeta^*=d/(\alpha-1)$ and a logarithmic singularity at $\zeta_s=-d$.  We shall not include the details here, but it is possible to derive a coupled system of odes for $a(t)$ and $d(t)$ that can be solved numerically or reduced further analytically by diving one by the other.  For $0<\alpha<2$ the singularity $\zeta_s$ is always smaller in magnitude, and therefore closer to the unit circle, than $\zeta^*$.  As a result, it turns out that $d$ is a decreasing function and that $\dot{d}\rightarrow 0^+$ as $d\rightarrow 1^+$.  Therefore, neither $\zeta^*$ nor $\zeta_s$ intersect the unit circle but in fact approach it asymptotically as $t\rightarrow\infty$.  As such, there is no cusp formation, and instead the proximity of $\zeta_s$ to the unit circle results in the interface forming a long finger, whose width is $(2-\alpha)\pi$.  In \Cref{fig:ZST}(e) we present an example with $\epsilon=0.2$, $\alpha=1.2$.  The interface in the top panel, given by the dashed red curves, clearly approaches a finger in shape.  The bottom panel shows the critical point (red dots) moving towards the unit circle, but always further away than the logarithmic singularity (black dots).

In each of \Cref{fig:ZST}(a)(top panel), (b)(top panel), (c)(left panel), (d)(top panel) and (e)(top panel), we have included numerical solutions drawn as solid blue curves that are computed using the same initial conditions but with a nonzero value of surface tension.  While we discuss these (more physically realistic) solutions at various points later in the paper, it is worth repeating here that nonzero surface tension is required in the model in order to relate to the real physics of a Hele-Shaw experiment.  The historical interest in complex singularities and finite-time blow-up is therefore mostly of a mathematical nature.

\section{Numerical scheme} \label{sec:NumericalScheme}

\subsection{The level set method}

To numerically solve \eqref{eq:HeleShaw1}-\eqref{eq:HeleShaw4}, following the methodology of \citet{Osher1988}, we construct a level set function $\phi$ such that the fluid-fluid interface $\partial \Omega$ is the zero level set of $\phi$ or
\begin{align}
\partial \Omega(t) = \left\lbrace \vec{x} \,| \, \phi(\boldsymbol{x}, t) = 0 \right\rbrace.
\end{align}
If the interface has the normal speed $v_n$, then we wish to construct a speed function, $F$, such that $v_n = F$ on $\boldsymbol{x} \in \partial \Omega(t)$, and is continuous over the entire computational domain. Thus $\phi$ satisfies the level set equation
\begin{align} \label{eq:LevelSetEquation}
\frac{\p \phi}{\p t} + F | \grad \phi | = 0.
\end{align}
We discuss how $F$ is computed in \Cref{sec:Speed}. To solve \eqref{eq:LevelSetEquation}, we approximate the spatial derivatives using a second order essentially non-oscillatory scheme (see \citet[chapter 3]{Osher2003} and \citet[chapter 6]{Sethian1999} for details), and integrate in time using second order total variation diminishing (TVD) Runge-Kutta (RK), which is performed by taking two forward Euler steps
\begin{align}
\tilde{\phi}^{(n + 1)} &= \phi^{(n)} - \Delta t F^{(n)} | \grad \phi^{(n)} |,\\
\phi^{(n + 2)} &= \tilde{\phi}^{(n+1)} - \Delta t F^{(n+1)} | \grad \tilde{\phi}^{(n+1)} |,
\end{align}
and then take an averaging step $\phi^{(n+1)} = \left( \phi^{(n)} + \phi^{(n + 2)} \right) / 2$. We note that the inclusion of the second order curvature term, $\kappa$, in the dynamic boundary condition \eqref{eq:HeleShaw20} would typically require $\Delta t \sim \Delta x^2$. However, we find that for the results presented in this work, the surface tension parameter is sufficiently small such that we can maintain numerical stability by choosing $\Delta t = \Delta x / (4 \max |F|)$.

The level set function $\phi$ is initialised as a signed distance function satisfying
\begin{align} \label{eq:LevelSetFunction}
	\phi = \begin{cases} d &\mbox{if } \boldsymbol{x} \in \mathbb{R}^2 \backslash \Omega(t) \\
		-d & \mbox{if } \boldsymbol{x} \in \Omega(t) \end{cases},
\end{align}
where $d$ is the minimum distance between $\boldsymbol{x}$ and $\partial \Omega$, via the method of crossing times \citep[chapter 7]{Osher2003}. That is, we advect $\phi$ in the normal direction to the interface by solving \eqref{eq:LevelSetEquation} with $F = 1$ and determine the point in time where each value of $\phi$ crosses from positive to negative. This process is repeated for $F = -1$. To reduce numerical error, which  can result when the gradient of $\phi$ becomes excessively small or large, we periodically perform re-initialisation in order to keep $\phi$ approximately equal to a signed distance function, which satisfies $|\grad \phi| = 1$, over the duration of a simulation. Re-initialisation is performed by solving
\begin{align} \label{eq:Reinit}
\frac{\p \phi}{\p \tau} + S(\phi) (|\grad \phi| - 1) = 0,
\end{align}
where
\begin{align}
S(\phi) = \frac{\phi}{\sqrt{\phi^2 + \Delta x^2}},
\end{align}
to steady state. Here $\tau$ is a pseudo time variable where $\Delta \tau = \Delta x / 5$. We find that performing re-initialisation every five time steps is sufficiently frequent.

While the level set method has successfully been used as a framework for studying a variety of moving boundary problems, a limitation of the method is that it can suffer from volume loss or gain in regions where the mesh is underresolved. In an effort to alleviate this problem, \citet{Enright2002} proposed the particle level set method, which combines the Eulerian based level set method with a marker particle based Lagrangian approach. We briefly describe the algorithm here, and refer the reader to \citet{Enright2002} for a more comprehensive description, as well as examples illustrating the effectiveness of the particle level set method.

The method works assigning placing massless particles in the regions where $\phi>0$ and $\phi<0$, i.e.~on both sides of the interface, which are referred to positive and negative particles, respectively. We denote $r_p$ as the minimum distance between the interface and particle's location The marker particles are advected according to
\begin{align}
	\frac{\textrm{d}\vec{x}_p}{\textrm{d} t} = F \vec{n} \label{eq:ParticleLSM},
\end{align}
where $\vec{x}_p$ is the location of the particle and $\vec{n} = \grad \phi / |\grad \phi|$ is a unit vector that reduces to the outward facing normal on the interface. If a particle crosses the interface, this indicates that mass has been lost (or gained). We mitigate this error by locally rebuilding the interface by constructing a local level set function from the four adjacent nodes to the particle defined as $\phi_p(\bmth{x}) = s_p(r_p - |\bmth{x} - \bmth{x}_p|)$, where $s_p = 1$ and $-1$ for positive and negative particles, respectively. Using $\phi_p$, $\phi$ is corrected using
$$\phi =
\begin{cases}
\phi^+ &\mbox{if } |\phi^+| \le |\phi^-| \\
\phi^- &\mbox{if } |\phi^+| > |\phi^-|
\end{cases},$$
where $\phi^+ = \max(\phi_p, \phi^+)$ and $\phi^- = \min(\phi_p, \phi^-)$. This procedure is performed both when the level set equation \eqref{eq:LevelSetEquation} is solved as well as when re-initialisation is performed.

\subsection{Solving for $F$} \label{sec:Speed}

\begin{figure}
	\centering
	\includegraphics[width=0.90\linewidth]{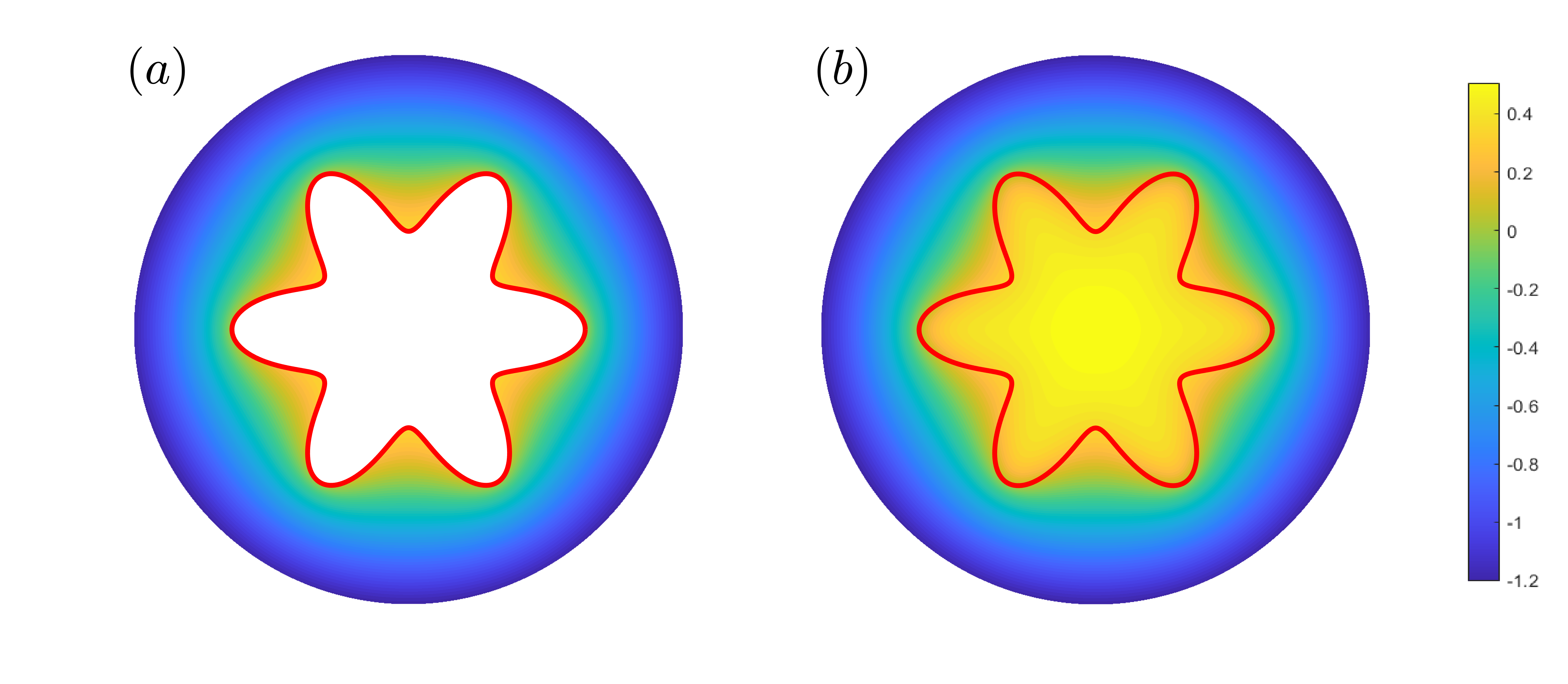}	
	\caption{An illustration of the velocity extension process used to extend $F$ to be defined over the entire computational domain. $(a)$ shows an example function that is undefined where $\bmth{x} \in \Omega$. $(b)$ shows this regions being `filled-in' by solving the biharmonic equation \eqref{eq:Biharmonic}. This gives us a differentiable extension of $F$ over the entire domain.}
	\label{fig:VelocityExtension}
\end{figure}

From the kinematic boundary condition \eqref{eq:HeleShaw3}, we have the expression for the speed function
\begin{align}
F =  -b^2 \grad p \cdot \boldsymbol{n}  \qquad \boldsymbol{x} \in \mathbb{R}^2 \backslash \Omega(t),	\label{eq:NormalSpeed}
\end{align}
recalling $F$ is required to solve the level set equation \eqref{eq:LevelSetEquation} and $\vec{n} = \grad \phi / |\grad \phi|$ reduces to the outward facing normal on the interface. Thus by solving \eqref{eq:NormalSpeed}, we find that $F = v_n$ on $\vec{x} \in \p \Omega(t)$, recalling $v_n$ is the normal speed of the interface. Further, \eqref{eq:NormalSpeed} provides a continuous expression for $F$ in the viscous fluid region. The derivatives in \eqref{eq:NormalSpeed} are evaluated using central differencing. However, to solve \eqref{eq:LevelSetEquation} we require an expression for $F$ over the entire computational domain. It was proposed by~\citet{Moroney2017} that the speed function be extended into the inviscid fluid region by solving the biharmonic equation
\begin{align} \label{eq:Biharmonic}
\nabla^4 F = 0  \qquad \boldsymbol{x} \in \Omega(t).
\end{align}
This ensures that $F = v_n$ is satisfied on the interface and gives a continuous expression for $F$ away from the interface.  For the purpose of discretisation, the sign of $\phi$ is used to determine nodes inside the interface that need to be included in the biharmonic stencil. This discretisation results in a symmetric system of linear equations that is solved using LU decomposition. As such, the location of the interface does not need to be known explicitly, similar to the level set method itself. This velocity extension process is a variant of a thin plate spline in two dimensions \citep{Bookstein1989}. To illustrate the velocity extension process, we consider an example $F$ defined in the region $\Omega(t)$, shown in \Cref{fig:VelocityExtension}$(a)$. The red line represents the fluid-fluid interface $\partial \Omega(t)$. \Cref{fig:VelocityExtension}$(b)$ shows $F$ after the biharmonic equation \eqref{eq:Biharmonic} is solved. We see that this velocity extension process gives a differentiable expression for $F$ over the entire computational domain, which can be used to solve \eqref{eq:LevelSetEquation}.

\subsection{Solving for pressure} \label{sec:SolvingPressure}

\begin{figure}
	\centering
	\includegraphics[width=0.45\linewidth]{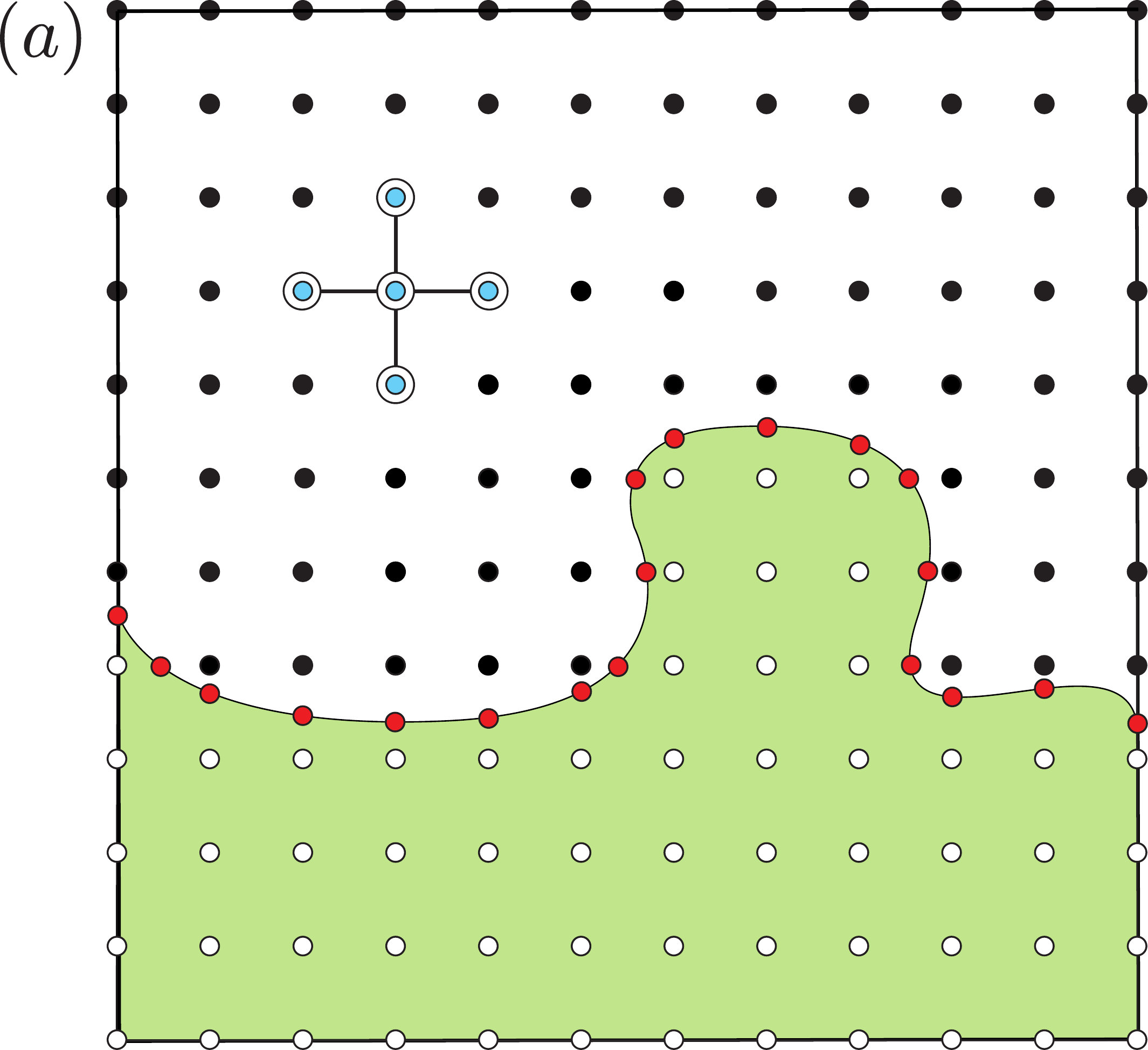}
	\includegraphics[width=0.45\linewidth]{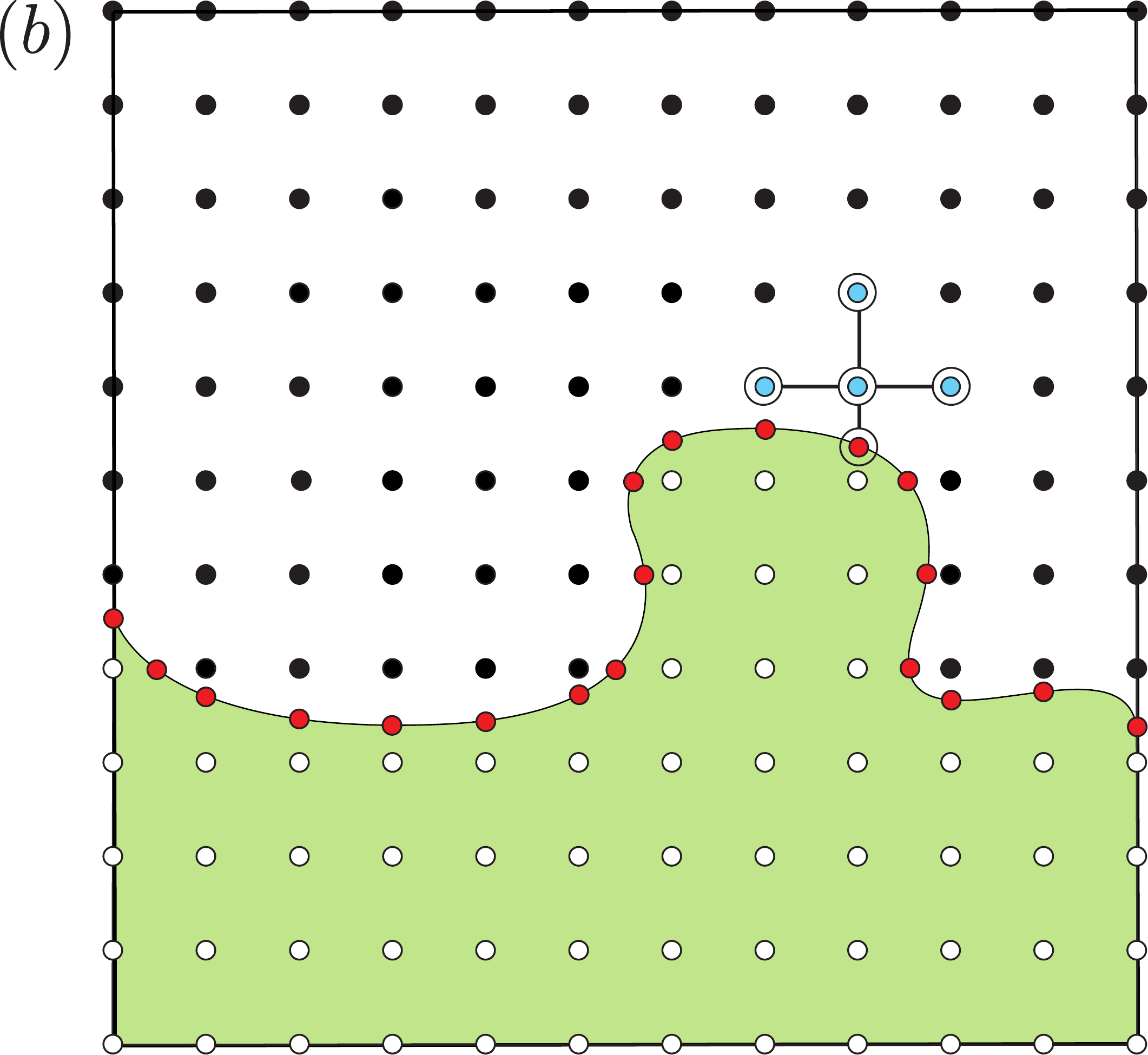}	
	\caption{An illustration of how the pressure of the viscous fluid is solved for using the finite difference method. $(a)$ For nodes away from the interface, we solve for pressure \eqref{eq:LaplacePolar} using a standard 5-point finite difference stencil \eqref{eq:FiniteDifference}. $(b)$ However, this stencil cannot be used for nodes adjacent to the interface, as in this case the southern node is not in the domain $\boldsymbol{x} \in \mathbb{R}^2 \backslash \Omega(t)$, and thus cannot be used in our stencil. Instead we impose a ghost node, denoted by the red dots, on the interface whose location corresponds to the point where $\phi=0$. The value at this ghost node is determined from the dynamic boundary condition \eqref{eq:HeleShaw2}. This leads to the non-standard finite difference stencil \eqref{eq:AdjustedFD}.}
	\label{fig:bc}
\end{figure}

To evaluate the speed function $F$, we must first compute the pressure field. We consider \eqref{eq:HeleShaw1}-\eqref{eq:HeleShaw4} in polar coordinates with $p = p(r, \theta, t)$ and the location of the interface is given by $r = s(\theta, t)$. Thus \eqref{eq:HeleShaw1} becomes
\begin{align} \label{eq:LaplacePolar}
\frac{1}{r} \frac{\partial}{\partial r}  \left( r b^3 \frac{\partial p}{\partial r} \right) + \frac{1}{r^2} \diffp{}{\theta} \left( b^3 \diffp{p}{\theta} \right) = \frac{\partial b}{\partial t} \qquad r > s(\theta, t).
\end{align}
For nodes away from the interface, the derivatives in \eqref{eq:LaplacePolar} are discretised using a standard 5-point stencil, illustrated in \Cref{fig:bc}$(a)$. Denoting $\beta = r b^3$, the $r$-derivatives in \eqref{eq:LaplacePolar} are approximated via
\begin{align}
\frac{1}{r} \frac{\partial}{\partial r} \left(  \beta \frac{\partial p}{\partial r} \right)  \to  \frac{1}{r_{i,j} \Delta r} \left(  \beta_{i+1/2,j} \frac{p_{i+1,j} - p_{i,j}}{\Delta r} - \beta_{i-1/2,j}  \frac{p_{i,j} - p_{i-1,j}}{\Delta r} \right), \label{eq:FiniteDifference}
\end{align}
where $\beta_{i+1/2,j} = (\beta_{i+1,j} + \beta_{i,j})/2$ and $\beta_{i-1/2,j} = (\beta_{i-1,j} + \beta_{i,j})/2$. The derivatives in the $\theta$-direction are discretised in a similar fashion.

As illustrated in \Cref{fig:bc}$(b)$, special care must be taken when solving for nodes adjacent to the interface. Suppose that the interface is located at $r = r_I$ where $r_{i-1,j} < r_I < r_{i,j}$ where the nodes $r_{i-1,j}$ and $r_{i,j}$ are in the inviscid and viscous fluid regions respectively. When discretising \eqref{eq:LaplacePolar}, we can no longer incorporate $p_{i-1,j}$ into our finite difference stencil as it is not in the domain $\boldsymbol{x} \in \mathbb{R}^2 \backslash \Omega(t)$. Instead, we define a ghost node at $r_I$ (denoted by the red dots in \Cref{fig:bc}$(b)$) whose value is $p_I$. By noting that $\phi$ is a signed distance function, the distance between $r_{i,j}$ and $r_I$ is computed via
\begin{align}
h = \Delta r \left| \frac{\phi_{i,j}}{\phi_{i-1,j} - \phi_{i,j}} \right|.
\end{align}
As per \citet{Chen1997}, our finite difference stencil becomes
\begin{equation} \label{eq:AdjustedFD}
\begin{split}
\frac{1}{r} \frac{\partial}{\partial r} \left( \beta \frac{\partial p}{\partial r} \right)  & \approx  \frac{2}{r_{i,j}(\Delta r + h)} \left( \beta_{i+1/2,j}  \frac{p_{i+1,j} - p_{i,j}}{\Delta r} -\hat{\beta}_{i-1/2,j} \frac{p_{i,j}}{h} \right) \\ &+ \frac{2}{r_{i,j} h (\Delta r + h)} \hat{\beta}_{i-1/2,j} p_I.
\end{split}
\end{equation}
Here $\hat{\beta}_{i-1/2,j} = (\beta_{i,j} + \beta_I)/2$ where $\beta_I$ is the value of $\beta$ on the interface. When the node and interface are sufficiently close together such that $h < \Delta r^2$, we set $p_{i,j} = p_I$. A similar procedure is applied if the interface lies between $r_{i,j} < r_I < r_{i+1,j}$ and in the azimuthal direction.

The value of $p_I$ is computed from the dynamic boundary condition \eqref{eq:HeleShaw2}. To determine the value of $p_I$, we first compute the curvature of $\phi$ via $\kappa = \grad \cdot \boldsymbol{n}$ over the entire computational domain, recalling $\boldsymbol{n} = \grad \phi / |\grad \phi|$. The value of $\kappa$ on the interface is computed via linear interpolation using $\kappa_{i,j}$ and $\kappa_{i-1,j}$ leading to
\begin{align}
p_I = -\sigma \left( \kappa_{i,j} - \frac{h(\kappa_{i,j}-\kappa_{i-1,j})}{\Delta r} + \frac{2}{b(r_I, \theta_j)} \right) - \omega^2 r_I^2.
\end{align}
For more information about solving both elliptic and parabolic problems in irregular domains using the finite difference method in conjunction with level set functions, we refer the reader to \citet{Coco2013,Gibou2002} (and references therein). Once the finite difference stencil has been formed, the resulting system of linear equations is solved using LU decomposition.

\subsubsection{Far-field boundary condition} \label{sec:Farfield}

To incorporate the far-field boundary condition \eqref{eq:HeleShaw4} into our finite difference stencil, we utilise a Dirichlet-to-Neumann map. This map is implemented by imposing an artificial circular boundary at $r = R$ such that $R > s(\theta, t)$. By only considering the region $s(\theta, t) \le r \le R$, we seek a solution to \eqref{eq:HeleShaw1} of the form
\begin{align} \label{eq:BC1}
\hat{p}(r, \theta, t) = A_0 - \frac{Q}{2 \pi} \log r + \frac{r^2}{4} \frac{\partial b}{\partial t} + \sum_{n=1}^{\infty} r^{-n} \left( A_n \cos n \theta + B_n \sin n \theta \right),
\end{align}
where $A_0$, $A_n$, and $B_n$ are unknown, and $\hat{p} = b^3 p$. The expansion \eqref{eq:BC1} assumes that $b$ is spatially uniform in $r \ge R$ so that $\hat{p}$ satisfies the appropriate Poisson equation. Considering the value of pressure on the artificial boundary, suppose that $\hat{p}(R, \theta, t)$ can be represented as a Fourier series
\begin{align} \label{eq:BC2}
\hat{p}(R, \theta, t) = a_0 + \sum_{n = 1}^{\infty} a_n \cos n \theta + b_n \sin n \theta,
\end{align}
where
$$
a_0 = \frac{1}{2\pi} \int_{0}^{2\pi} \hat{p}(R, \theta,t) \hspace{0.15em} \text{d} \theta,
\quad
a_n = \frac{1}{\pi} \int_{0}^{2\pi} \hat{p}(R, \theta,t) \cos n \theta \hspace{0.15em} \text{d}\theta,
\quad
b_n = \frac{1}{\pi} \int_{0}^{2\pi} \hat{p}(R, \theta,t) \sin n \theta \hspace{0.15em}  \text{d}\theta.
$$
By equating \eqref{eq:BC2} with \eqref{eq:BC1} evaluated at $r = R$, we find that $A_0 = a_0 + (Q/2 \pi) \log R - \dot{b}R^2 / 4$, $A_n = R^n a_n$ and $B_n = R^n b_n$.

To incorporate our expression for $\hat{p}$ into our finite difference stencil, we differentiate \eqref{eq:BC2} with respect to $r$ and evaluate it at $r = R$ to give
\begin{align}
\frac{\partial}{\partial r} \hat{p}(R, \theta_j) = -\frac{Q}{2 \pi R} + \frac{R}{2} \frac{\partial b}{\partial t} - \sum_{n = 1}^{\infty} \frac{n}{R} \left(  a_n \cos n \theta_j + b_n \sin n \theta_j \right).
\end{align}
By approximating the integrals in our expressions for $a_n$ and $b_n$ as
\begin{align}
a_n \approx \frac{\Delta \theta}{\pi} \sum_{k=1}^{m} \hat{p}(R, \theta_k) \cos n \theta_k \quad \textrm{and} \quad b_n \approx \frac{\Delta \theta}{\pi} \sum_{k=1}^{m} \hat{p}(R, \theta_k) \sin n \theta_k,
\end{align}
then
\begin{align}
\frac{\partial}{\partial r} \hat{p}(R, \theta_j)\approx -\frac{Q}{2 \pi R} + \frac{R}{2}\frac{\partial b}{\partial t} - \frac{\Delta \theta}{R \pi} \sum_{k = 1}^{m}  v_{jk} \hat{p}(R, \theta_k),
\end{align}
where
\begin{align}
v_{jk} = \sum_{n = 1}^{\infty} n \cos (n (\theta_k - \theta_j)).
\end{align}
Defining $I$ as the outermost index at which $r = R$, then our expression for $\partial p / \partial r$ is incorporated into our finite difference stencil,
\begin{dmath}
	\frac{1}{r} \frac{\partial}{\partial r} \left(  \beta \frac{\partial p}{\partial r} \right) \to  \frac{2}{R \Delta r}   \left\lbrace -\beta_{I-1/2,j} \frac{  p_{I,j} - p_{I-1,j}}{\Delta r} +  R  \left[ -\frac{Q}{2 \pi R} + \frac{R}{2} \frac{\partial b}{\partial t} - b^3 \frac{\Delta \theta}{R \pi} \sum_{k = 1}^{m} w_{jk} p_{I,k}   \right]   \right\rbrace,
\end{dmath}
\noindent recalling $\beta = r b^3$.

As an aside, we note this procedure can be adapted to model a Dirichlet boundary condition of the form $p \sim p_{\infty}$ as $r \to \infty$ where $p_\infty$ is prescribed. This type of boundary condition would be more appropriate for the model of a Stefan problem \citep{Morrow2019b}, where now $p$ would represent temperature that is prescribed in the far-field. Assuming that $b$ is constant, then \eqref{eq:BC1} becomes
\begin{align}
p = p_\infty + A_0\ln r + \sum_{n=1}^{\infty} r^{-n} \left( A_n \cos n \theta + B_n \sin n \theta \right),
\end{align}
By following the same steps outlined above, we have
\begin{align}
\frac{\partial}{\partial r} p(R, \theta_j) &= \frac{a_0 - p_\infty}{R \ln R} - \sum_{n = 1}^{\infty} \frac{n}{R} \left(  a_n \cos n \theta_j + b_n \sin n \theta_j \right).
\end{align}
This expression is then incorporated into our finite difference stencil as per usual.

\subsection{General algorithm}

We summarise our numerical algorithm as follows:

\begin{itemize}
	\item[\textit{Step 1}] For a given initial interface $s(\theta, 0)$, initialise $\phi$ as a signed distanced function such that it satisfies \eqref{eq:LevelSetFunction} using the method of crossing times.
	\item[\textit{Step 2}] Place marker particles around the interface, noting which side of the interface they are on, as well as their minimum distance from the interface.
	\item[\textit{Step 3}] Solve for pressure in the domain $\vec{x} \in \mathbb{R}^2 \backslash \Omega$ using the procedure described in \Cref{sec:SolvingPressure}.
	\item[\textit{Step 4}] Compute $F$ according to \eqref{eq:NormalSpeed}, and then extend $F$ into the region $\vec{x} \in \Omega$ by solving the biharmonic equation \eqref{eq:Biharmonic}.
	\item[\textit{Step 5}] Using $F$, update both $\phi$ and the marker particles by solving \eqref{eq:LevelSetEquation} and \eqref{eq:ParticleLSM}, respectively.
	\item[\textit{Step 6}] Correct $\phi$ (if necessary) using the marker particles.
	\item[\textit{Step 7}] Re-initialise the level set function by solving \eqref{eq:Reinit}, and then correct $\phi$ using the marker particles.
	\item[\textit{Step 8}] Repeat steps 2-7 until $t = t_f$.
\end{itemize}

\section{Numerical results}

In this section, we present a selection of results demonstrating the capabilities of the numerical scheme presented in \Cref{sec:NumericalScheme}. We show how our framework can be modified to solve for a wide range of different configurations of the Hele--Shaw cell, and provide examples illustrating that simulations are capable of producing solutions consistent with previous experimental and numerical results.

\subsection{Expanding bubble problem} \label{sec:Standard}

We first consider the standard Hele--Shaw problem in which the inviscid bubble is injected into the viscous fluid while the plates are parallel and stationary such that $b = 1$ and $\omega = 0$.

\subsubsection{Numerical validation} \label{sec:Verification}

\begin{figure}
	\centering
	\includegraphics[width=0.35\linewidth]{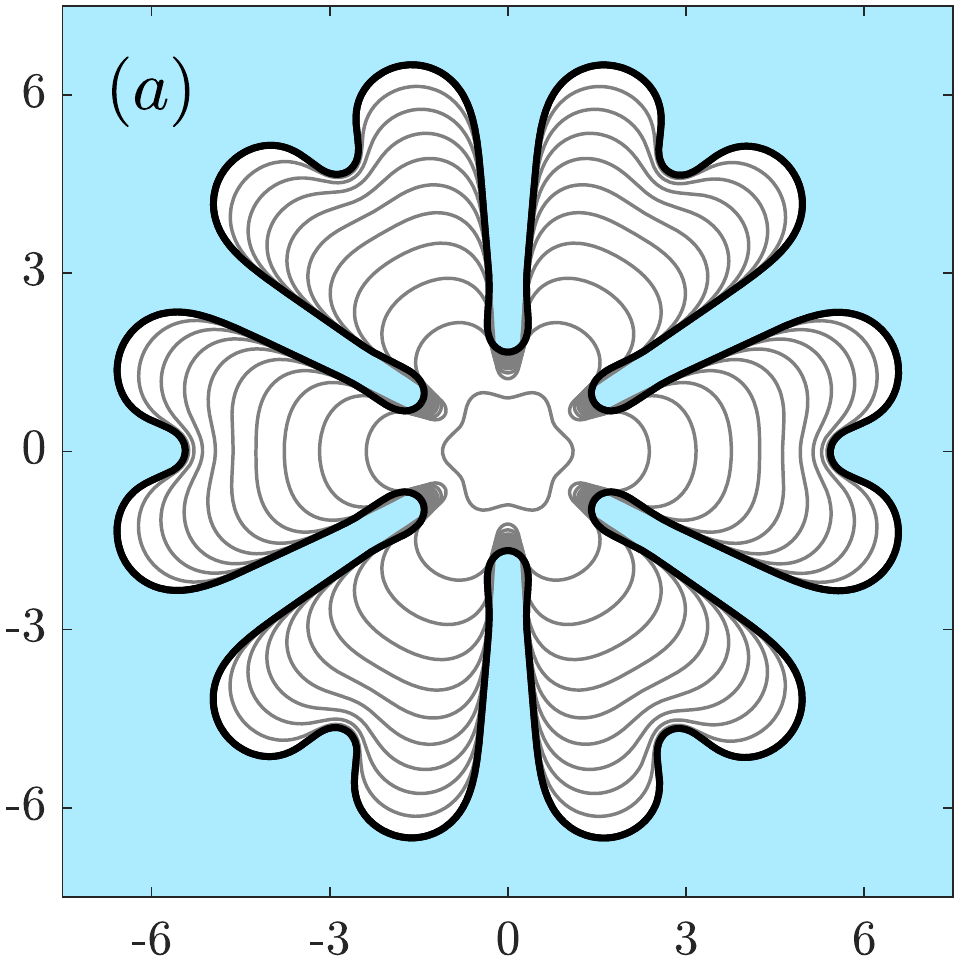} \hspace{1em}
	\includegraphics[width=0.35\linewidth]{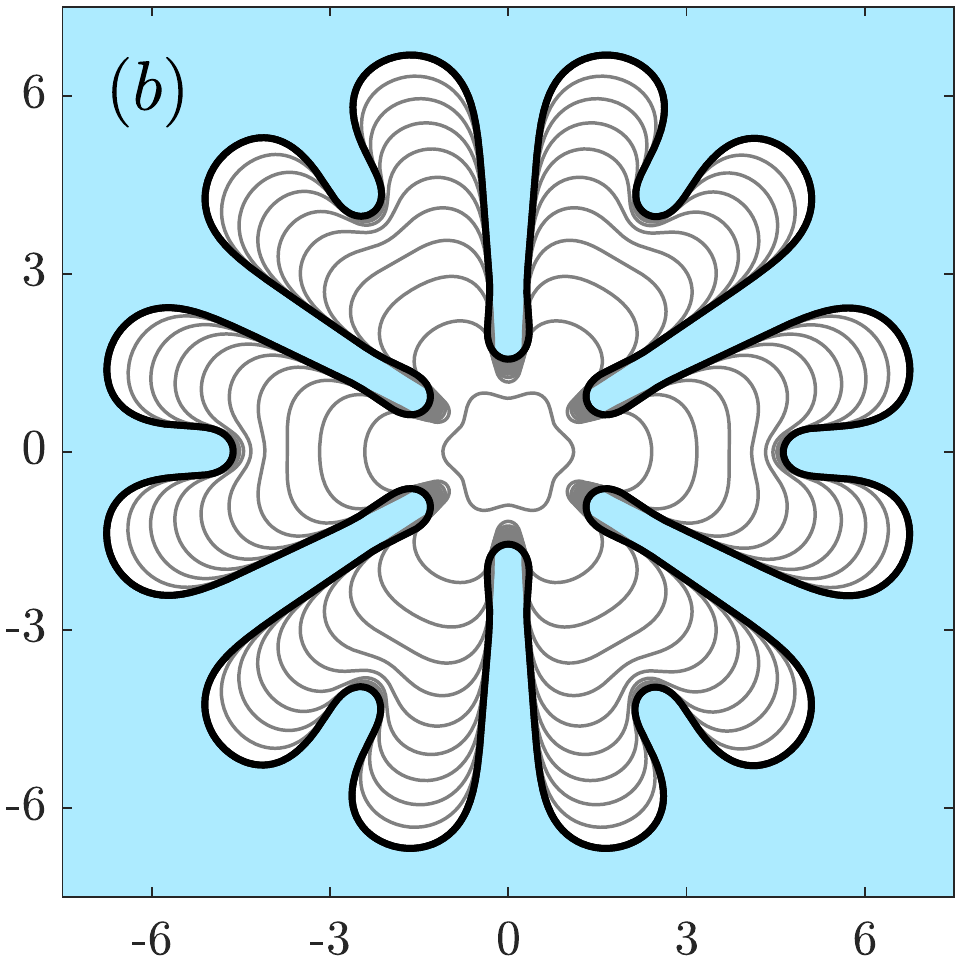}\\
	\vspace{1em}
	\includegraphics[width=0.35\linewidth]{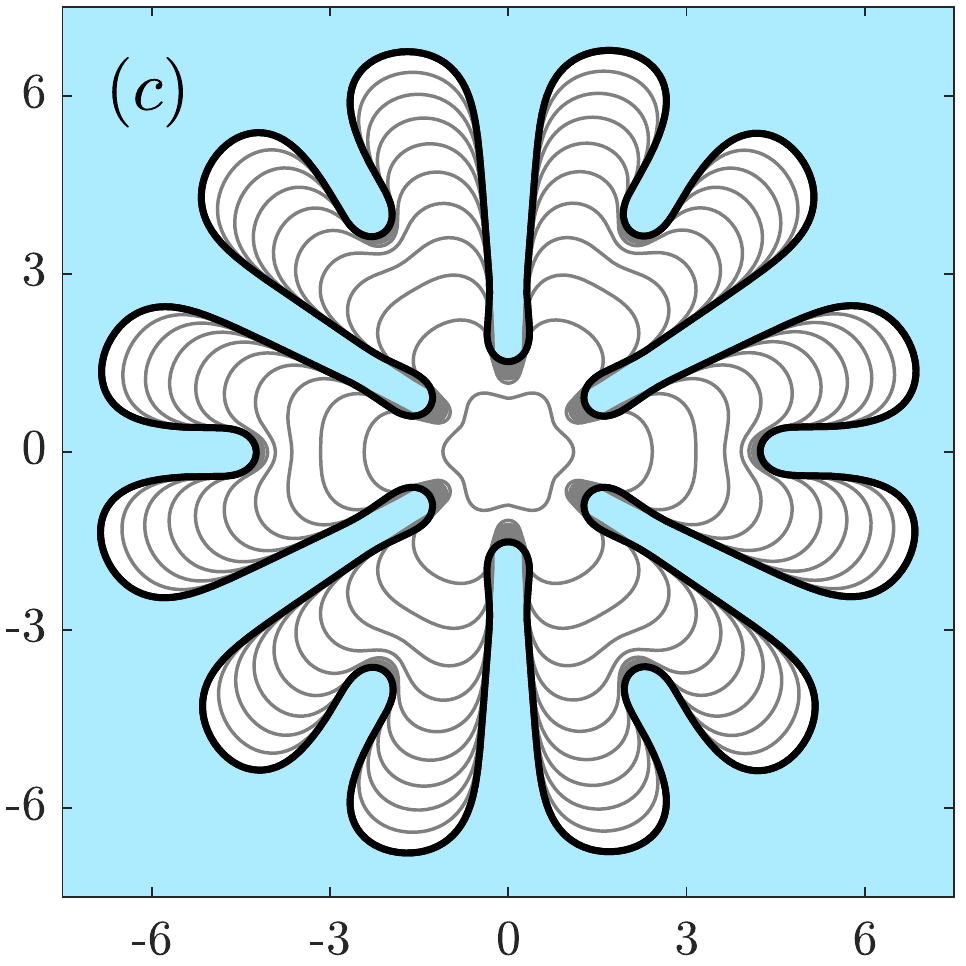} \hspace{1em}
	\includegraphics[width=0.35\linewidth]{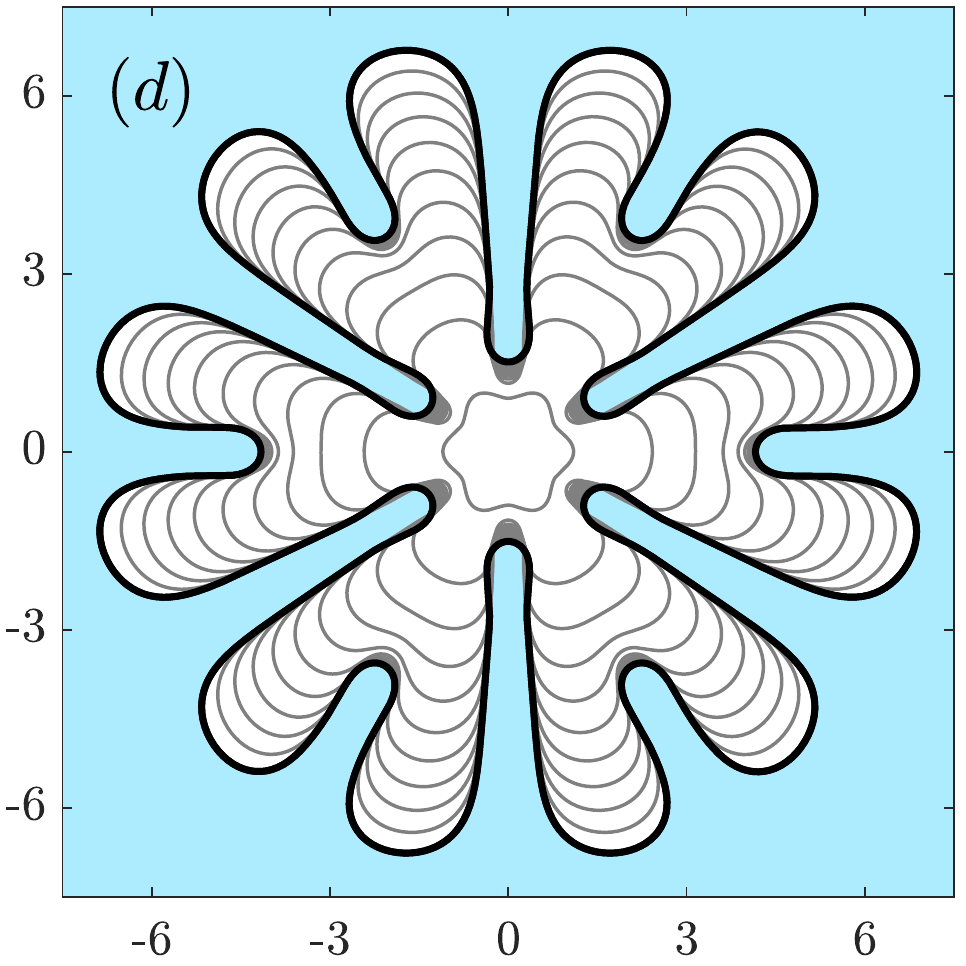}			
	\caption{Convergence test of numerical scheme for the evolution of a bubble with initial condition \eqref{eq:GridRefinementIC}, where solutions are computed using $(a)$ $350 \times 293$, $(b)$ $550 \times 461$, $(c)$ $750 \times 628$, and $(d)$ $850 \times 712$ equally spaced nodes. Additionally, $Q = 1$, $\sigma = 5 \times 10^{-4}$, $\omega = 0$, and $t_f = 100$. Simulations are performed on the domain $0 \le r \le 7.5$ and $0 \le \theta < 2\pi$. Solutions are plotted in time intervals of $t = 10$.}
	\label{fig:GridRefinement}
\end{figure}

As a preliminary test for our scheme, we demonstrate that numerical solutions converge for a sufficiently refined grid. To do so, we perform simulations with the initial condition
\begin{align} \label{eq:GridRefinementIC}
s(\theta, 0) = 1 + \varepsilon \cos m \theta,
\end{align}
where $\varepsilon = 0.1$, $m = 6$, on the computational domain $0 \le r \le 7.5$ and $0 \le \theta < 2 \pi$, and are performed using an increasingly refined mesh. These simulations, shown in \Cref{fig:GridRefinement}, indicate that the interfacial profiles are converging as the mesh is refined, and that grid independence is achieved, at this scale, using $750 \times 628$ equally spaced nodes. Further, our solutions appear to maintain six fold symmetry over the duration of the simulation.

\begin{figure}
	\centering
	\includegraphics[width=0.45\linewidth]{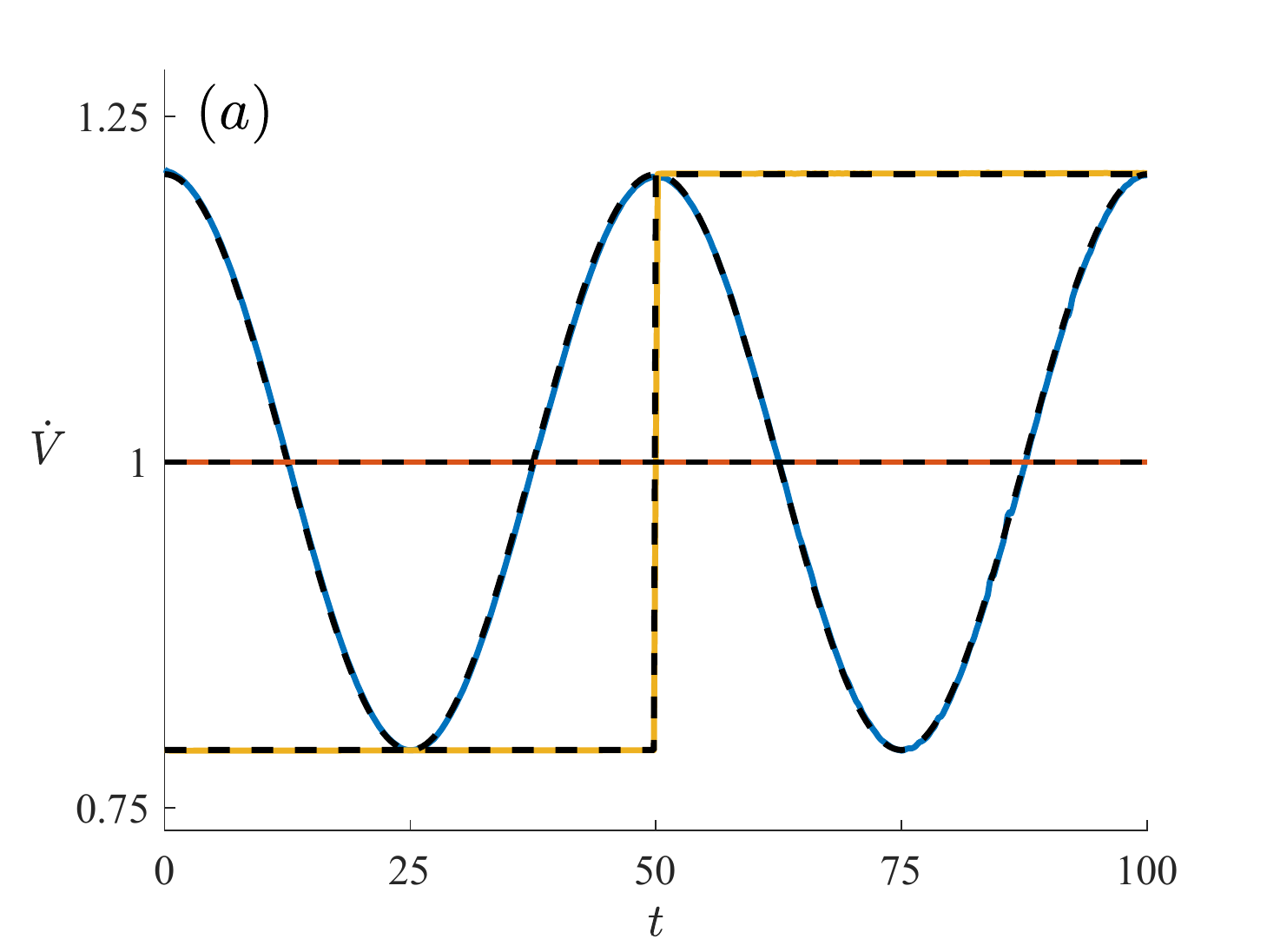}
	\includegraphics[width=0.45\linewidth]{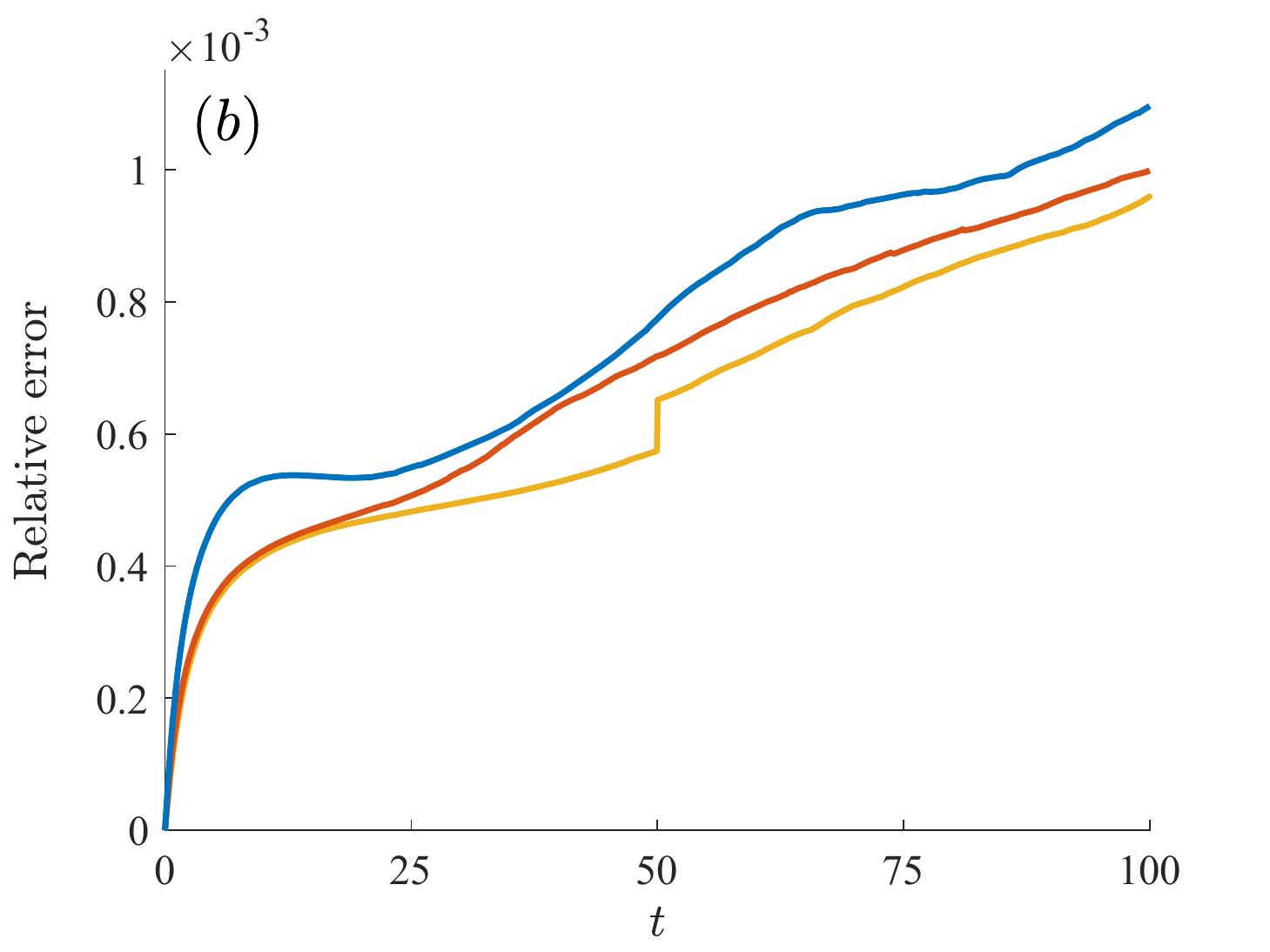}
	\caption{$(a)$ The rate of change of volume, $\dot{V}$, of numerical solution with constant (red), periodic (blue) and piecewise (yellow) injection rates. Dotted black lines denote the corresponding exact rate of change of volume. $(b)$ Corresponding relative error. Simulations are performed on the domain $0 \le r \le 7.5$ and $0 \le \theta < 2 \pi$ using $750 \times 628$ equally spaces nodes with the initial condition \eqref{eq:GridRefinementIC}. The surface tension parameter is $\sigma = 5 \times 10^{-4}$ and final time of simulations is $t = 100$.}
	\label{fig:timedependentinjection}
\end{figure}

We also demonstrate that our use of the Dirichlet-to-Neumann map, described in \Cref{sec:Farfield}, results in the bubble's volume changing at rate $Q$. To do so, we consider three different injection rates. The first is the constant injection rate $Q = 1$, the second is the sinusoidal injection rate,
\begin{align}
Q =  1 + 0.2 \sin (4\pi t/t_f),
\end{align}
and the third is the piecewise rate
\begin{align}
Q = \left\{
\begin{array}{ll}
0.8 & \quad t \leq t_f/2 \\
1.2 & \quad t > t_f/2
\end{array}.
\right .
\end{align}
We find that the rate of change of volume computed from the numerical simulations compares well with the corresponding exact rate of expansion, shown in \Cref{fig:timedependentinjection}$(a)$. The relative error, shown in \Cref{fig:timedependentinjection}$(b)$, suggests that over the duration of a simulation, we experience mass loss of only approximately $0.1\%$. This result suggests that the Dirichlet-to-Neumann map correctly ensures that the bubble expands at the correct rate for both constant and time-dependent $Q$.

\subsubsection{Effect of surface tension} \label{eq:SurfaceTension}

Surface tension, modelled via the inclusion of the signed curvature term in the dynamic boundary condition \eqref{eq:HeleShaw2}, acts to regularise the Hele--Shaw model.  As shown in \Cref{sec:complex}, in the absence of surface tension, exact solutions exist that exhibit very different behaviour (finite-time cusp formation or finger formation) despite initial conditions that are arbitrarily close; hence the zero-surface-tension-problem is ill-posed. Adding surface tension removes the possibility of cusp formation; in \Cref{fig:ZST}, we have included solutions for small but nonzero surface tension, computed using the method described in \Cref{sec:NumericalScheme}, for each of the cases described in \Cref{sec:complex}. With nonzero surface tension, the interface remains smooth and solutions exist for all time, or, in the case of the finite blob (\Cref{fig:ZST}(c)), until the interface intersects the point at which fluid is being withdrawn.

\begin{figure}
	\centering
	\includegraphics[width=0.4\linewidth]{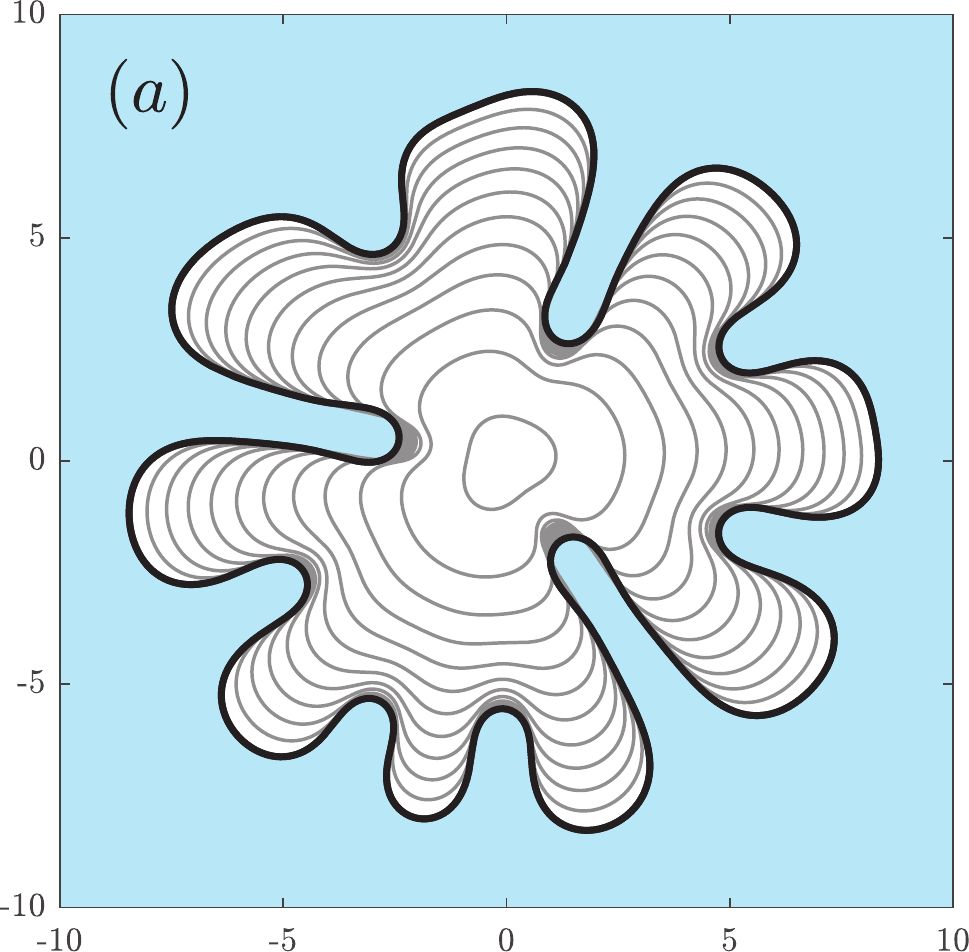} \hspace{1em}
	\includegraphics[width=0.4\linewidth]{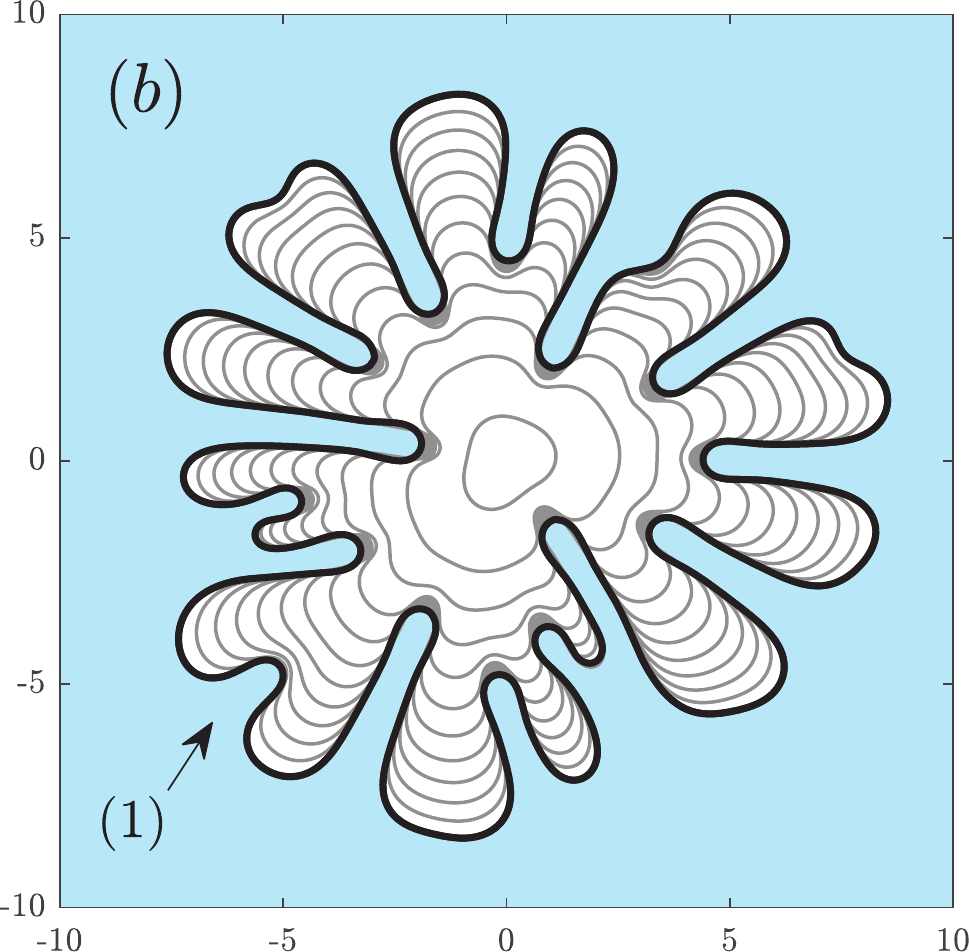} \\
	\vspace{1em}
	\includegraphics[width=0.4\linewidth]{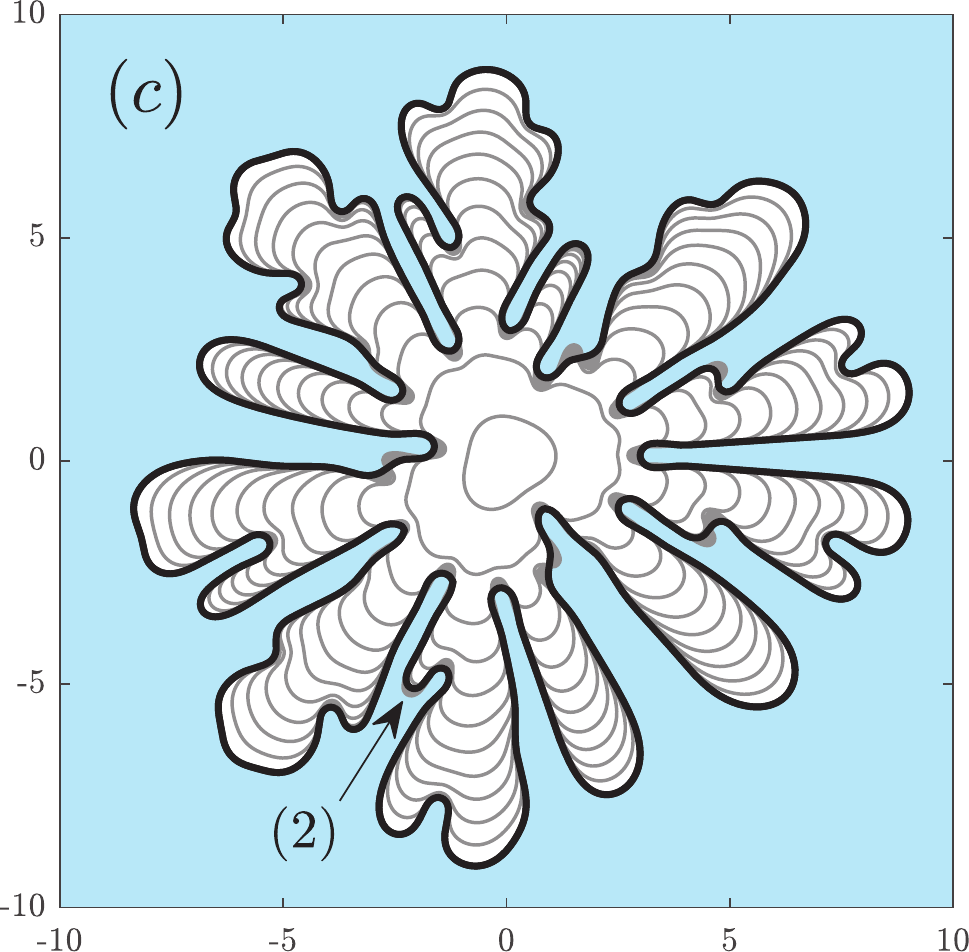}	\hspace{1em}
	\includegraphics[width=0.4\linewidth]{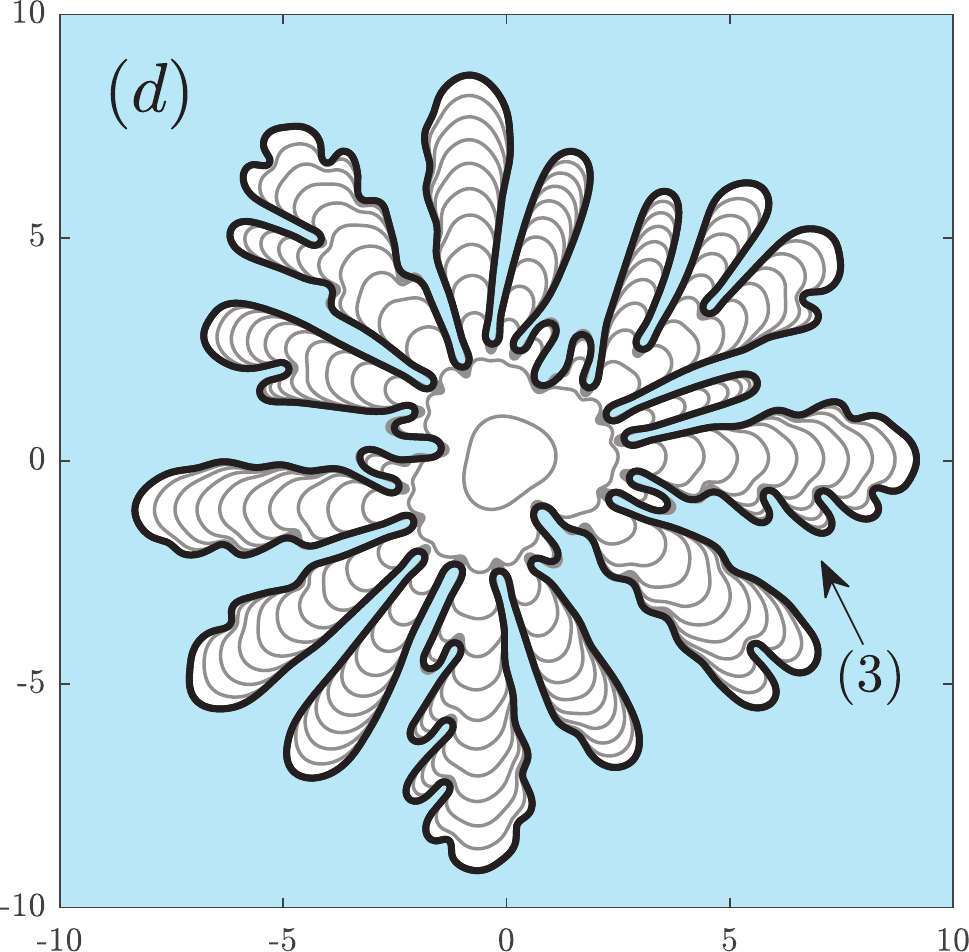}		
	\caption{The numerical solution to \eqref{eq:HeleShaw1}-\eqref{eq:HeleShaw4} with $Q = 1$ for values of surface tension parameter $\sigma$ $(a)$ $1.5 \times 10^{-3}$, $(b)$ $5 \times 10^{-4}$, $(c)$ $1.75 \times 10^{-4}$, and $(d)$ $6.25 \times 10^{-5}$ with initial condition \eqref{eq:InitialCondition}. Our numerical scheme captures different morphological behaviour including $(1)$ tip-splitting, $(2)$ shielding, and $(3)$ feathering, all of which have been observed experimentally. Simulations are performed on the domain $0 \le r \le 10$ and $0 \le \theta < 2 \pi$ using $1000 \times 628$ equally spaced nodes.	Solutions are plotted in time intervals of $t = 5$ up to $t_f = 55$.}
	\label{fig:SurfaceTension}
\end{figure}

Linear stability analysis \citep{Paterson1981} indicates that increasing the surface tension parameter $\sigma$ acts to make the interface less unstable, and nonlinear numerical simulations and experiments \citep{Chen1987,Dai1993,Hou1997} show that increasing the injection rate, which is mathematically equivalent to decreasing the dimensionless surface tension $\sigma$, results in an increase in the number of fingers that develop. We perform numerical simulations for $Q=1$ with values of surface tension varying over several orders of magnitude with the initial condition
\begin{align} \label{eq:InitialCondition}
s(\theta, 0) = 1 + \varepsilon \left( \cos m \theta + \sin n \theta \right),
\end{align}
where $\varepsilon = 0.1$, $m = 3$, and $n = 2$. These numerical simulations, shown in \Cref{fig:SurfaceTension}, are able to reproduce the key morphological features of the Saffman--Taylor instability. For each value of $\sigma$ considered, the interface is unstable, and the sinusoidal perturbations in the initial condition \eqref{eq:InitialCondition} grow and evolve into viscous fingers. As $\sigma$ is decreased, the interface becomes more unstable and the number of fingers that develop over the duration of a simulation increases due to tip-splitting occurring (see $(1)$ in \Cref{fig:SurfaceTension}$(b)$ for example). Additionally, our simulations are able to produce so called `shielding' behaviour, where neighbouring fingers can block one another off, which in turn results in fingers retracting (denoted by $(2)$ in \Cref{fig:SurfaceTension}$(c)$). This behaviour is known to occur experimentally (see Figure 1 in \citep{Paterson1981} for example). Finally, when surface tension is sufficiently small, `feathering' can occur, when a finger does not strictly tip-split, but instead develops ripples along one of its sides as it expands (denoted by $(3)$ in \Cref{fig:SurfaceTension}$(d)$). Again this behaviour has been observed in experiments (see Figure 3 in \citep{Chen1987}).

\subsection{Tapered plates} \label{sec:TaperedPlatesRadial}

One of the more popular modifications to the Hele--Shaw cell (particularly in recent years) is to consider that the plates of the Hele--Shaw cell are no longer parallel but instead linearly tapered (either converging or diverging) in the direction of the flow. The concept was first introduced in \citet{Zhao1992}, who considered tapered plates in a channel geometry, discussed further in \Cref{sec:Channel}. Imposing tapered plates in radial geometry has been studied using linear stability analysis \citep{Al2012}, weakly-nonlinear stability analysis \citep{Anjos2018}, experimentally \citep{Al2012,Bongrand2018}, and by numerical simulations \citep{Jackson2017}. Results indicate that tapering the gap between the plates either such that they converge or diverge can have a stabilising or de-stabilising effect on the interface depending on the injection rate.   The numerical scheme presented in \Cref{sec:NumericalScheme} was used by us in Ref.~\citet{Morrow2019} to study how tapering the plates while injecting at either a constant or time dependent injection rate can be used to reduce the development of viscous fingering patterns.

\begin{figure}
	\centering
	\includegraphics[width=0.40\linewidth]{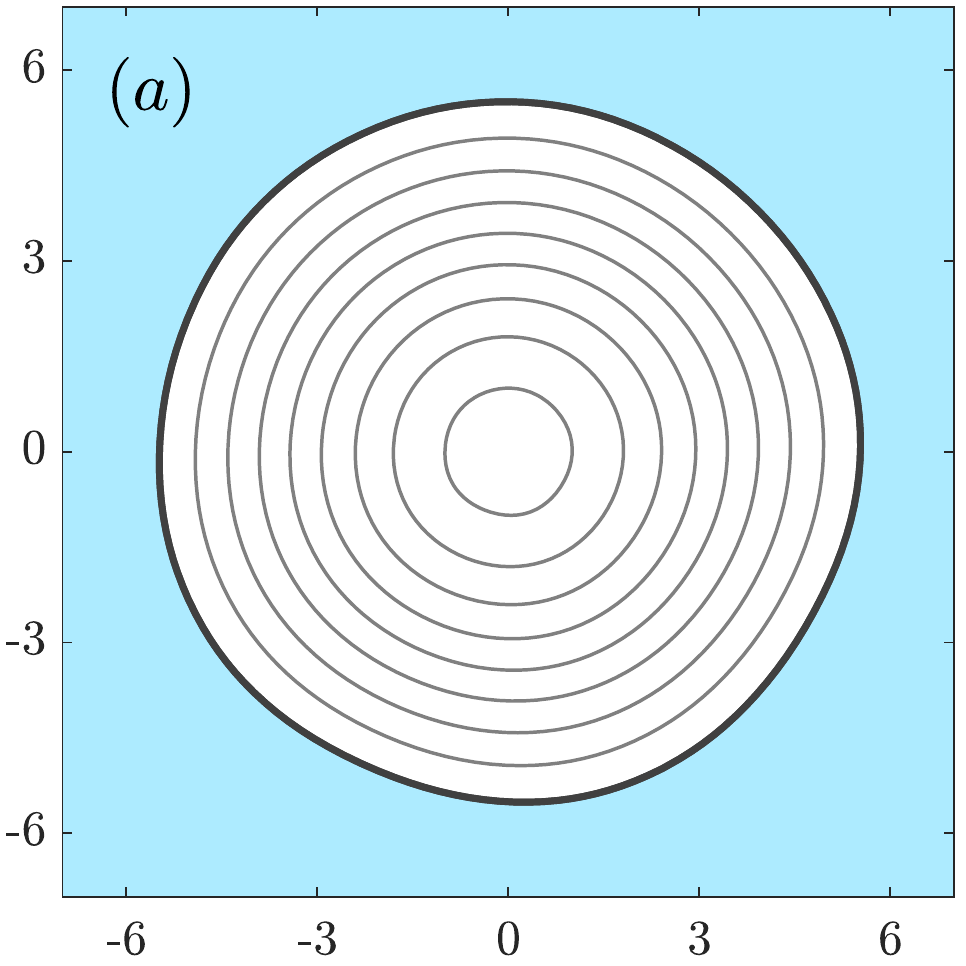} \hspace{1em}
	\includegraphics[width=0.40\linewidth]{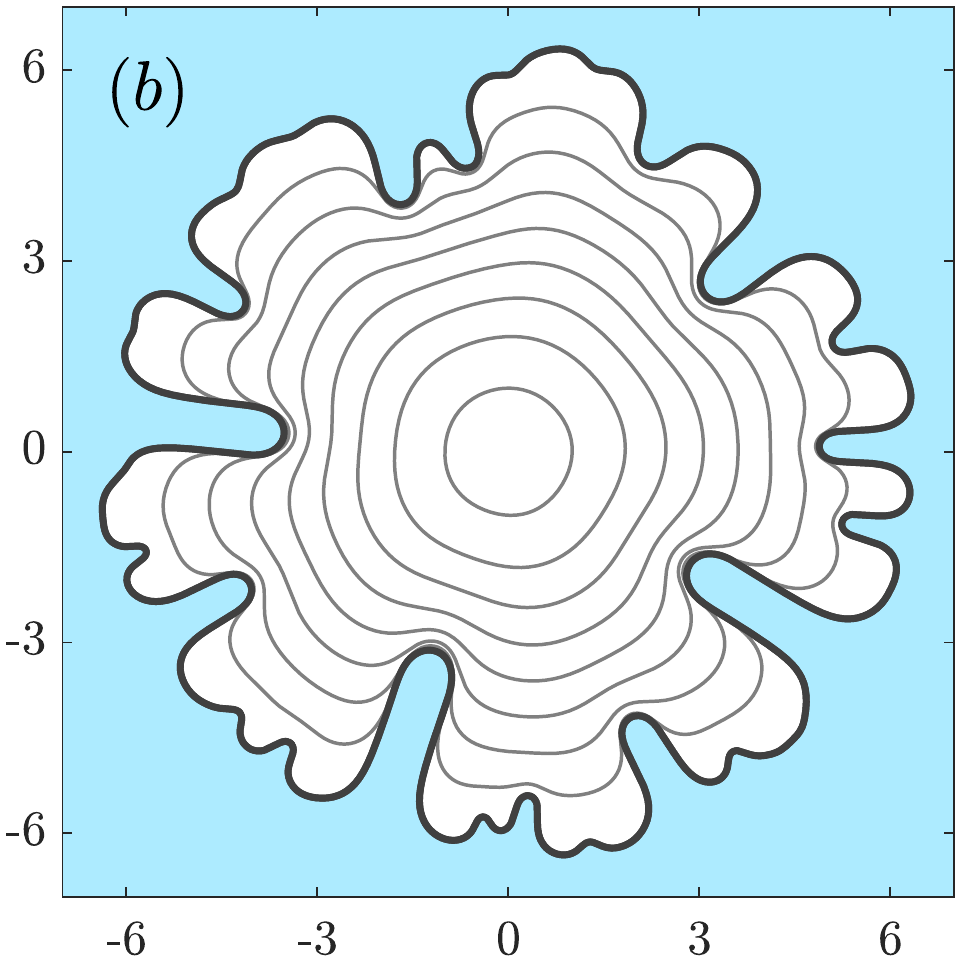} \\	
	\caption{Numerical simulations where the gap between the plates linearly tapered according to \eqref{eq:TaperedPlate} with $Q = 1$, $b_\infty = 1/15$, $R_0 = 8/3$, $r_B = 7$, and $\alpha = 2/15$, with surface tension $\sigma$ $(a)$ 6 and $(b)$ 1. Simulations are performed on the domain $0 \le r \le 7.5$ and $0 \le \theta < 2 \pi$ using $750 \times 628$ equally spaced nodes. Solutions are plotted in time intervals of 5.6. up to $t_f = 44.8$.}
	\label{fig:TaperedPlates}
\end{figure}

Here we perform numerical simulations where the gap between the plates is linearly tapered according to
\begin{align} \label{eq:TaperedPlate}
b(r) =
\begin{cases}
b_\infty + \alpha (r - r_B) & \text{if } r \le r_B, \\
b_\infty     & \text{if } r > r_B,
\end{cases}
\end{align}
where $\alpha$ is the gradient of the taper ($\alpha=0$ being the unmodified Hele--Shaw cell). Experiments were recently performed by \citet{Bongrand2018}, who considered a Hele--Shaw cell where the plate gap is \eqref{eq:TaperedPlate} where $\alpha > 0$ for different injection rates. Bongrand and Tsai showed that if the injection rate is sufficiently small, the interface is completely stabilised over the duration of the experiment.  In our nondimensionalisation, for which the time-scale is set by the injection rate, a smaller (dimensional) injection rate corresponds to a larger (dimensionless) surface tension value $\sigma$.  To confirm that our numerical scheme produces solutions consistent with these experiments, we perform simulations with two different values of $\sigma$, shown in \Cref{fig:TaperedPlates}, to model two different (dimensional) injection rates.  The initial condition of these simulations is (\ref{eq:InitialCondition}) where $0 \le \theta < 2 \pi$, $\varepsilon = 5 \times 10^{-3}$, $m = 5$, and $n = 4$. For a larger $\sigma$ (\Cref{fig:TaperedPlates}$(a)$), we indeed find that the interface is stabilised, and remains circular over the duration of the simulation. For a smaller $\sigma$ (\Cref{fig:TaperedPlates}$(b)$), we find that the interface is unstable; in particular, these fingers appear distinct from traditional Saffman--Taylor fingers (see \Cref{fig:SurfaceTension} for example). This behaviour is consistent with the results of \citet{Bongrand2018}, who described the interface as becoming `wavy' as it expanded. These results suggest our numerical simulations are producing the correct behaviour when the plates are of the form of \eqref{eq:TaperedPlate}.

\subsection{Blob problem} \label{sec:Complementary}

In this subsection, we now consider the complementary problem to \eqref{eq:HeleShaw1}-\eqref{eq:HeleShaw4}, for which a blob of viscous fluid now occupies the region $\Omega(t)$ and the inviscid fluid is in $\mathbb{R}^2 \backslash \Omega(t)$.  This problem is typically studied by considering the withdrawal of the viscous fluid, which in turn causes fingers to develop inward (as in \Cref{fig:ZST}(c), for example). However, popular modifications, including considering the gap between the plates to be time-dependent or rotating the entire Hele--Shaw cell, have also received interest. In this subsection, we explain how the numerical scheme presented in \Cref{sec:NumericalScheme} is modified to solve for these variations, and show that our simulations are consistent with previous experimental and numerical results.

For the case in which the inviscid bubble is injected into the viscous fluid (\Cref{sec:Standard}-\ref{sec:TaperedPlatesRadial}, say), the velocity of the fluid is driven by a sink term in the far-field given by \eqref{eq:HeleShaw4}. For the blob problem considered here, the withdrawal of the viscous fluid is incorporated into the model via a sink at the origin. To include this sink into our numerical model, we follow \citet{Hou1997} and introduce a smoothed Dirac delta function in \eqref{eq:HeleShaw1} such that
\begin{align} \label{eq:HeleShaw5}
\grad \cdot \left( b^3 \grad p \right) = \frac{\p b}{\p t} + S,
\end{align}
where
\begin{align} \label{eq:Sink}
S = \begin{cases}
\dfrac{Q}{b \bar{r}_0^2} \left( 1 + \cos \dfrac{\pi r}{\bar{r}_0} \right) & \text{if } r \leq \bar{r}_0, \\
0       & \text{if } r > \bar{r}_0.
\end{cases}
\end{align}
As shown by \citet{Hou1997}, this choice of source term ensures the rate of change of volume of the viscous fluid is $Q$. For all results in this subsection, we use $\bar{r}_0 = 0.05$. We note that it is straightforward to extend \eqref{eq:Sink} to include multiple sink/source points.

When solving for the governing equation for pressure \eqref{eq:HeleShaw1} for the case where in inviscid fluid is injected into a viscous one (\Cref{sec:Standard}-\ref{sec:TaperedPlatesRadial}), we consider the model in polar coordinates as it is simpler to incorporate the far-field boundary condition \eqref{eq:HeleShaw4} via the Dirichlet-to-Neuman map (\Cref{sec:Farfield}) on a circle. However, when the inviscid region surrounds the viscous fluid, we no longer have a far-field boundary condition and, as such, it is more convenient to solve for pressure in Cartesian coordinates, although of course either coordinate system could be used. The discretisation of the spatial derivatives \eqref{eq:HeleShaw5} is performed as was described in \Cref{sec:SolvingPressure}. That is, we use central finite differences for nodes away from the interfaces, and incorporating a ghost value of $p$ for nodes adjacent to the interface. We refer to Refs~\citet{Chen1997,Gibou2002,Gibou2003} for more details on implementing this stencil in Cartesian coordinates.

\begin{figure}
	\centering
	\includegraphics[width=0.90\linewidth]{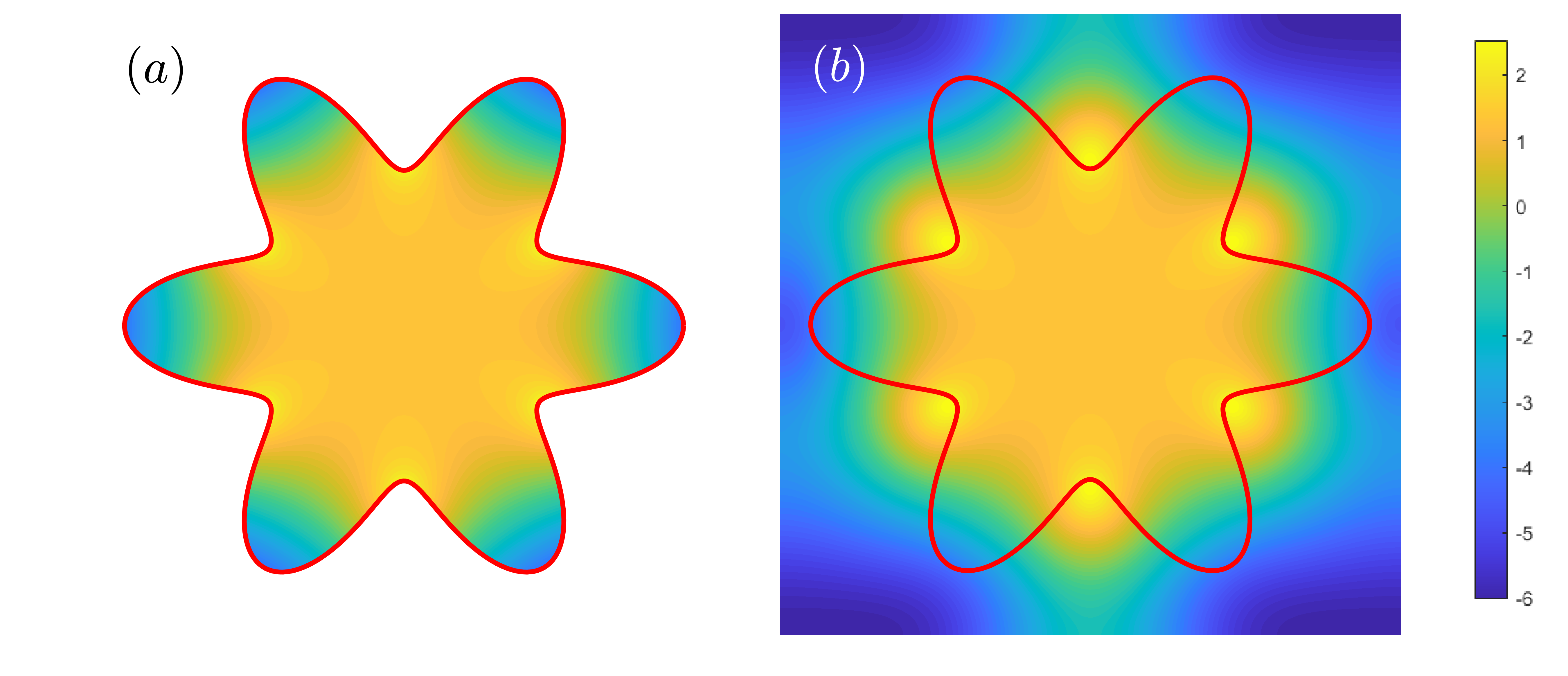}	
	\caption{An illustration of the velocity extension process where the viscous fluid is surrounded by the inviscid bubble. As discussed in \Cref{sec:Speed}, this is done by solving the biharmonic equation in the region where $\boldsymbol{x} \in \Omega(t)$, where now we apply homogeneous Neumann boundary conditions on the edge of the computational domain.}
	\label{fig:VelocityExtension2}
\end{figure}

In a similar way to that described in \Cref{sec:Speed}, once we have computed $F$ via \eqref{eq:NormalSpeed}, we extend it into the region $\boldsymbol{x} \in \mathbb R^2 \backslash \Omega(t)$ by solving the biharmonic equation \eqref{eq:Biharmonic}. When solving \eqref{eq:HeleShaw1}-\eqref{eq:HeleShaw4}, the speed function was known outside the interface and was extended in. We now have an expression for $F$ inside the interface that needs to be extended outward. This means we require boundary conditions on each of the four computational boundaries. When forming our biharmonic stencil, we apply homogeneous Neumann boundary conditions on each of the boundaries. We illustrate this process in \Cref{fig:VelocityExtension2}, which shows an example $F$ before (\Cref{fig:VelocityExtension2}$(a)$) and after (\Cref{fig:VelocityExtension2}$(b)$) the velocity extension process. Again we see that by solving the biharmonic equation we obtain a smooth expression for $F$ over the entire computational domain.

\subsubsection{Withdrawal of viscous fluid}

We consider the case where the viscous fluid is withdrawn such that as the interface contracts, viscous fingers form inward.  As described in \Cref{sec:complex}, exact solutions in the absence of surface tension ($\sigma = 0$) are ill-posed, and unphysical cusps can form before the interface reaches the sink (the point at which liquid is withdrawn). Numerical simulations performed using the boundary integral method have investigated the regularising effects of surface tension preventing these cusps from forming \citep{Ceniceros1999,Kelly1997}.  Experimental results \citep{Paterson1981,Thome1989} show that the fingers that form exhibit morphological features distinct from traditional Saffman--Taylor fingers, in that these fingers do not appear to undergo tip-splitting but instead appear to be in competition to `race' toward the sink.

\begin{figure}
	\centering
	\includegraphics[width=0.45\linewidth]{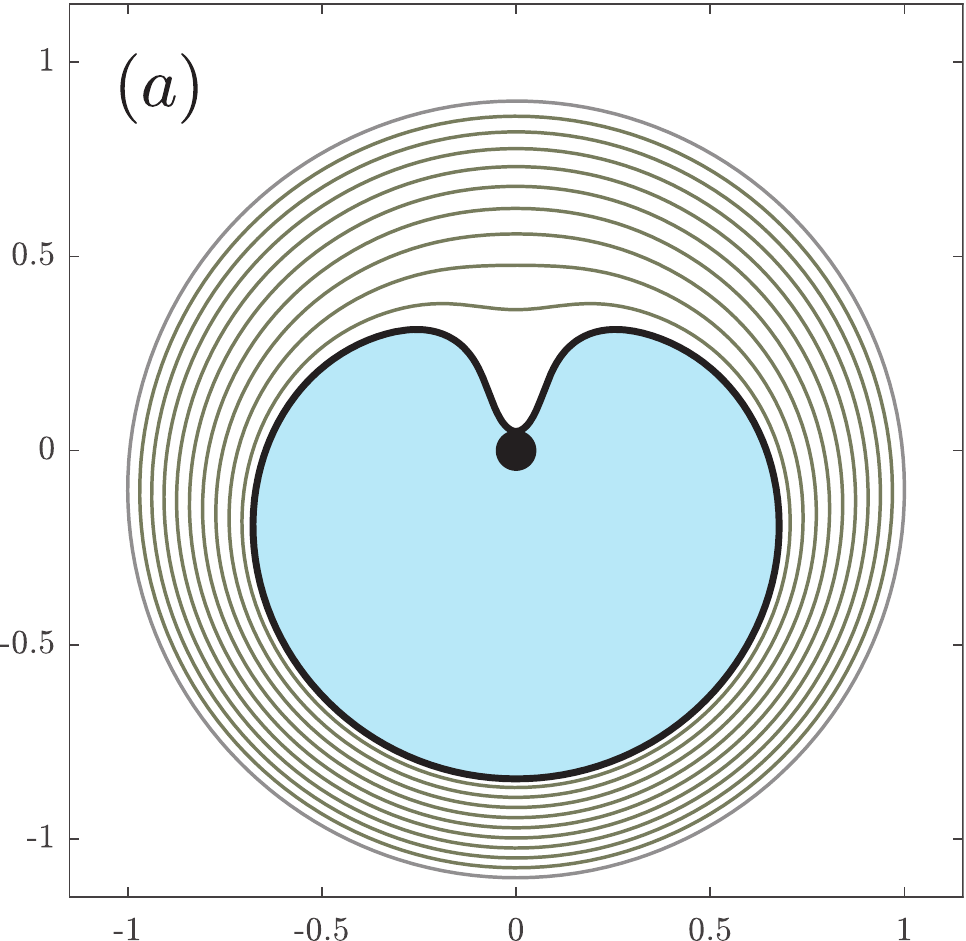} \hspace{1em}
	\includegraphics[width=0.45\linewidth]{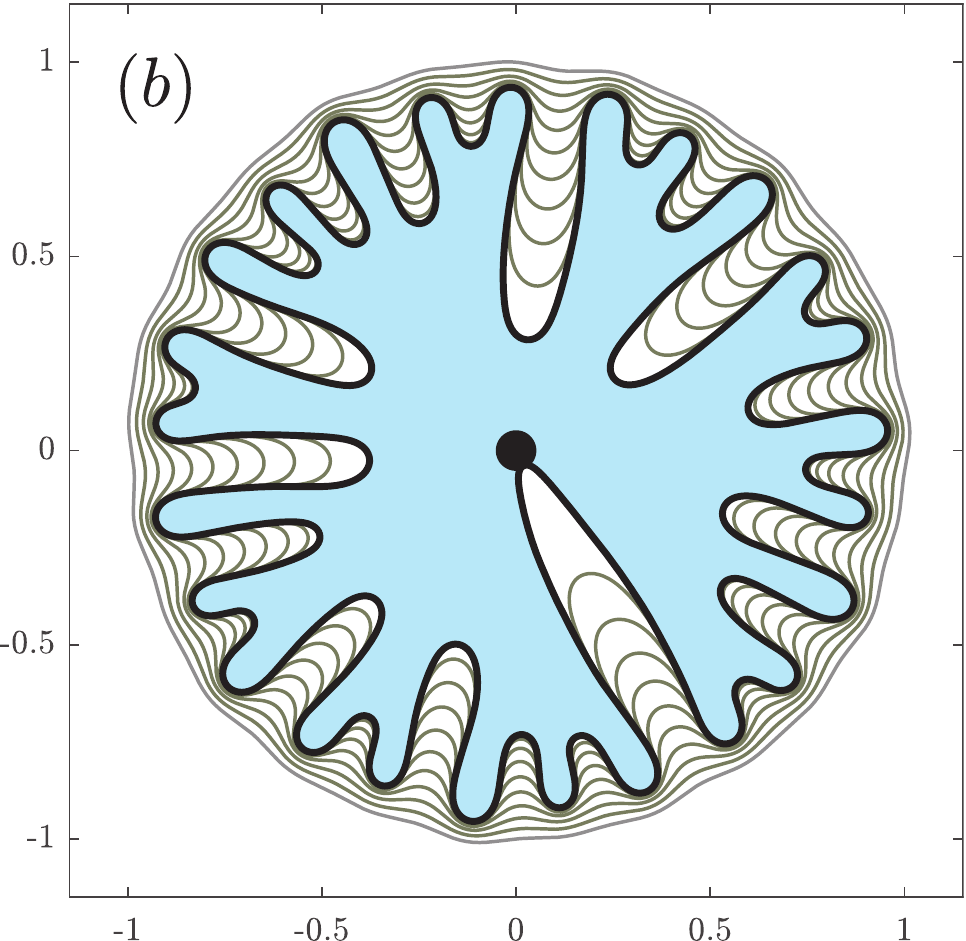}
	\caption{Numerical simulation where the viscous fluid is withdrawn at a point located at the origin where $Q = 1$. For $(a)$, initial condition is a circle of radius unity centred at $(0.1, 0)$ where $\sigma = 8 \times 10^{-4}$ and $t_f = 1.88$. For $(b)$, initial condition is \eqref{eq:InitialCondition2} where $\sigma = 1.6 \times 10^{-6}$ and $t_f = 1.45$. Black dot denotes region where $r \le 0.05$. Simulations are performed on the domain $-1.15 \le x \le 1.15$ and $-1.15 \le y \le 1.15$ using $400 \times 400$ equally spaced nodes.}
	\label{fig:Suction}
\end{figure}

As well as the comparison with the zero surface tension solution made in \Cref{fig:ZST}(c), we perform two further different simulations for this configuration. The first is with an initially circular interface centred at $(0,-0.1)$ shown in \Cref{fig:Suction}$(a)$. This simulation shows that as the interface contracts, it becomes non-convex until a single finger develops that tends towards the origin.  This behaviour compares well with previous numerical simulations performed using the boundary integral method \citep{Kelly1997}. For the second simulation, shown in \Cref{fig:Suction}$(b)$, we consider a perturbed circle centred at the origin of the form
\begin{align} \label{eq:InitialCondition2}
s(\theta, 0) = 1 + \varepsilon \left( \cos 3 \theta + \sin 7 \theta + \cos 15 \theta + \sin 25 \theta \right),
\end{align}
where $\varepsilon = 5 \times 10^{-3}$. We find that the interface initially develops numerous short fingers. These fingers do not appear to exhibit the same morphological features as the case in which the inviscid bubble is injected such as tip-splitting and feathering (see \Cref{fig:SurfaceTension} for example), but instead the number of fingers remains constant. Due to the pressure differential between the sink and the boundary of the bubble, the velocity of one of the fingers rapidly increases, and the simulation is stopped when this finger reaches the origin. We note that this behaviour compares well with experimental results (see Figure 15 in \citep{Thome1989} for example), as well as numerical simulations in \citep{Chen2014}.

\subsubsection{Lifting plates} \label{sec:LiftingPlates}

A popular modification to the blob problem is to consider the case where the upper plate is uniformly lifted in time such that $b \to b(t)$. The volume of viscous fluid remains constant ($Q = 0$) so that when the plates are separated, viscous fingers developing inward.  For this problem, the Hele-Shaw approximation itself can only remain valid for as long as the gap $b(t)$ is sufficiently small.  For example, in dimensional terms, we must be very careful about using the model when the gap width is of the same order as the important length scales in the lateral direction.

For this lifting plates configuration, the governing equation for pressure \eqref{eq:HeleShaw5} becomes Poisson's equation. However, pressure can be reduced via $P = p - \dot{b} |\boldsymbol{x}|^2/(4 b^3)$, essentially moving the non-homogeneous term to the dynamic boundary condition \eqref{eq:HeleShaw2}. This transformation reduces \eqref{eq:HeleShaw1} to Laplace's equation, meaning this configuration can also be solved numerically with the boundary integral method \citep{Zhao2018,Zhao2021}.

The lifting plate problem was first considered mathematically by \citet{Shelley1997}. In the absence of surface tension ($\sigma = 0$), \citet{Shelley1997} argued that the generic behaviour is that a cusp will develop in a finite time.  When surface tension is included, numerical simulations suggested a relationship between the number of fingers that develop and the surface tension parameter. In particular, it was shown that the number of fingers is a monotonically decreasing function of time. This behaviour was later shown to be consistent with experimental results \citep{Lindner2005,Nase2011}. As a point of comparison, for the traditional Hele--Shaw configuration, discussed in \Cref{sec:Standard}, the number of fingers typically increases with time due to tip-splitting. We note that the case in which the inviscid bubble is injected into the viscous fluid while the plates are separated has also received attention \citep{Morrow2019,Vaquero2019,Zheng2015}.

\begin{figure}
	\centering
	$\sigma = 10^{-4}$ \\
	\includegraphics[width=1.0\linewidth]{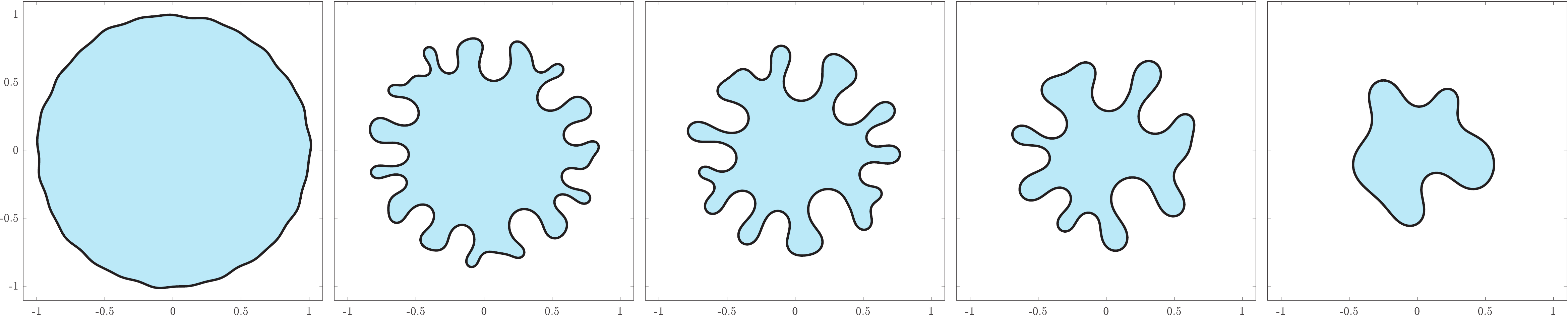}\\
	\vspace{0.25cm}
	$\sigma = 5 \times 10^{-5}$ \\
	\includegraphics[width=1.0\linewidth]{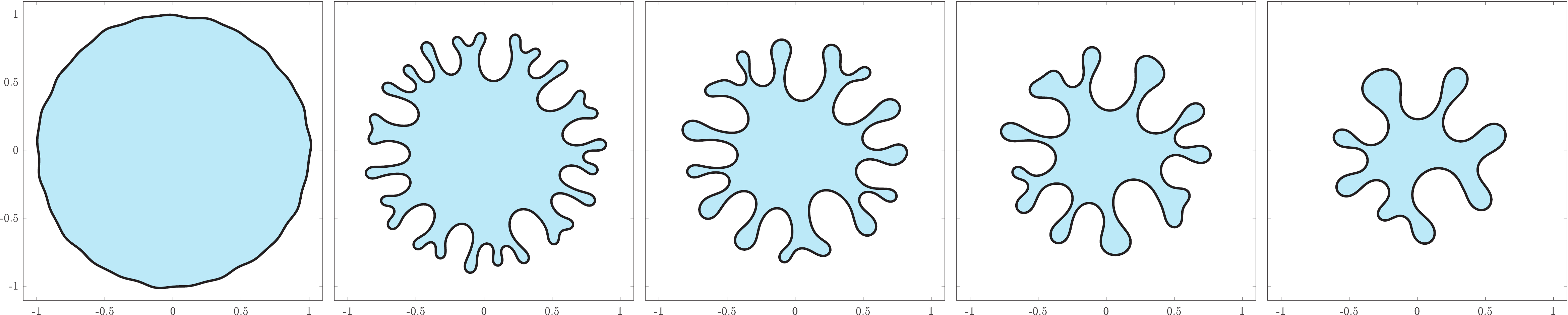}	
	\caption{Time evolution of viscous fluid where plates are separated according to $b = 1 + t$ with initial condition \eqref{eq:InitialCondition2}. Solutions are plotted at times (left to right) $t = 0$, 1, 1.8, 2.8, and 4. Simulations are performed on the domain $-1.1 \le x \le 1.1$ and $-1.1 \le y \le 1.1$ using $400 \times 400$ equally spaced nodes.}
	\label{fig:LiftingPlates}
\end{figure}

We perform a simulation using the linearly increasing gap between the plates $b = 1 + t$ for different values of $\sigma$. Note that this time-dependence with $\dot{b}=1$ effectively chooses the appropriate time-scale $T$ in (\ref{eq:scaledvariables}). When $\sigma = 10^{-4}$ (row one of \Cref{fig:LiftingPlates}), we find that the interface quickly destabilises and approximately 15 fingers develop by $t = 1$. As time increases further, neighbouring fingers begin to merge with each other and the overall number of fingers significantly decreases, and we find only around five fingers remain by the conclusion of the simulation. For the lower value of surface tension $\sigma = 5 \times 10^{-5}$ (row two of \Cref{fig:LiftingPlates}), we find the number of fingers that initially develop has increased compared to when $\sigma = 10^{-4}$, about 25 at $t = 1$. However as the interface contracts, again fingers begin to merge and only 8 fingers remain at $t = 4$. We perform these simulations over a longer time period (not shown), which reveals that the interface will become circular when the gap between the plates is sufficiently large. This behaviour is consistent both with previous experimental and numerical results \citep{Lindner2005,Nase2011,Shelley1997}.

\subsubsection{Rotating plates} \label{sec:RotatingPlates}

While Saffman--Taylor fingers traditionally form due to the injection/withdrawal of one immiscible fluid into another, it is known that these fingers can also be triggered by body forces. The two most commonly studied body forces are gravity and centrifugal forces. For the latter, when the entire Hele--Shaw cell is rotated, this rotation results in the dense viscous fluid being propelled outward, which in turn leads to  finger formation \citep{Schwartz1989}.  Experimental \citep{Carrillo1996} and numerical \citep{Alvarez2008,Folch2009,Paiva2019} studies reveal that the interface patterns are distinct from the traditional Saffman--Taylor instability. That is, fingers appear more `stretched-out' and generally go not undergo tip-splitting. We note that our model \eqref{eq:HeleShaw1}-\eqref{eq:HeleShaw5} ignores the effect of Coriolis forces, however several studies have investigated its effect on interfacial dynamics \citep{Alvarez2008,Schwartz1989,Waters2005}. Further, we consider the case where the inviscid bubble is injected into the viscous fluid while the plates are rotated in \citet{Morrow2019}, where we show that the angular velocity has a stabilising effect on the interface.

\begin{figure}
	\centering
	\includegraphics[width=0.4\linewidth]{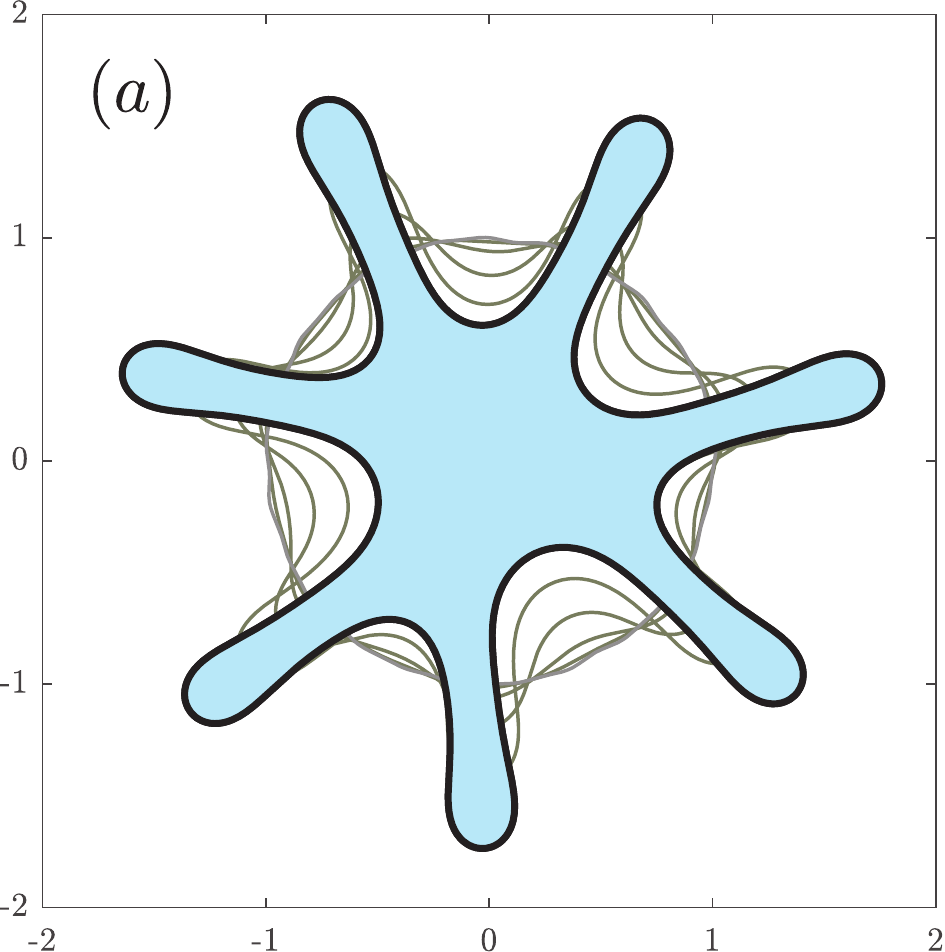} \hspace{1em}
	\includegraphics[width=0.4\linewidth]{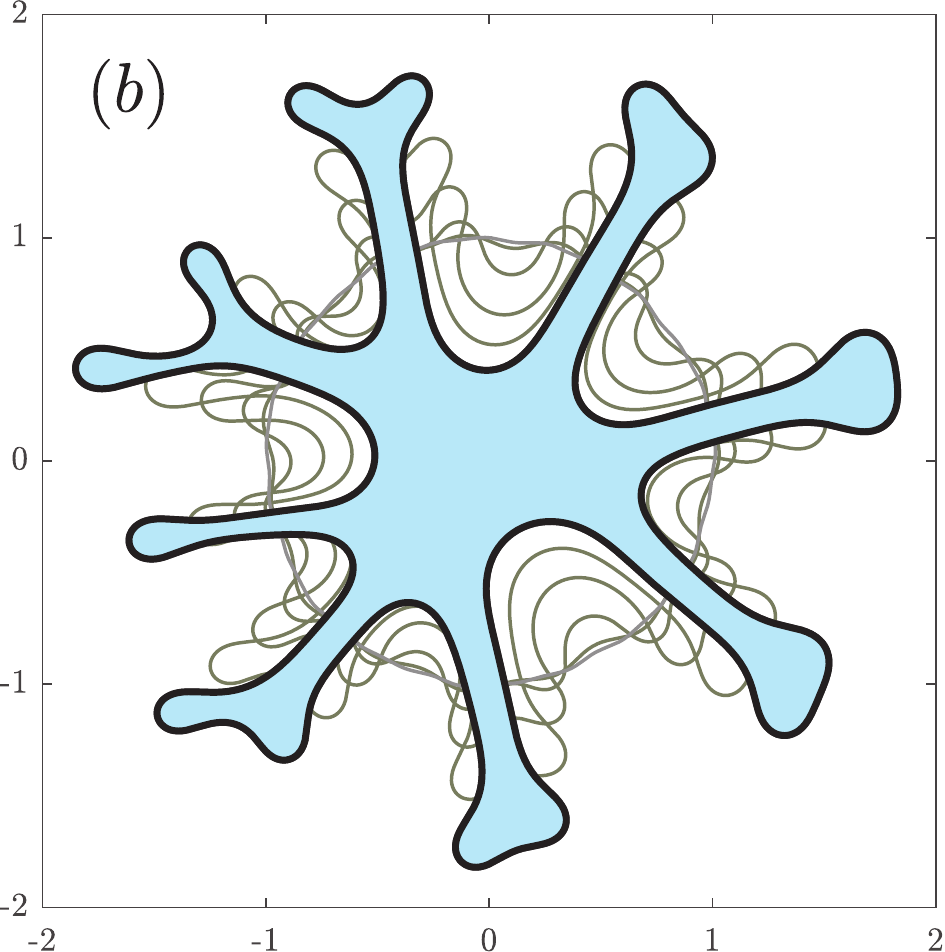} \\
	\vspace{1em}
	\includegraphics[width=0.4\linewidth]{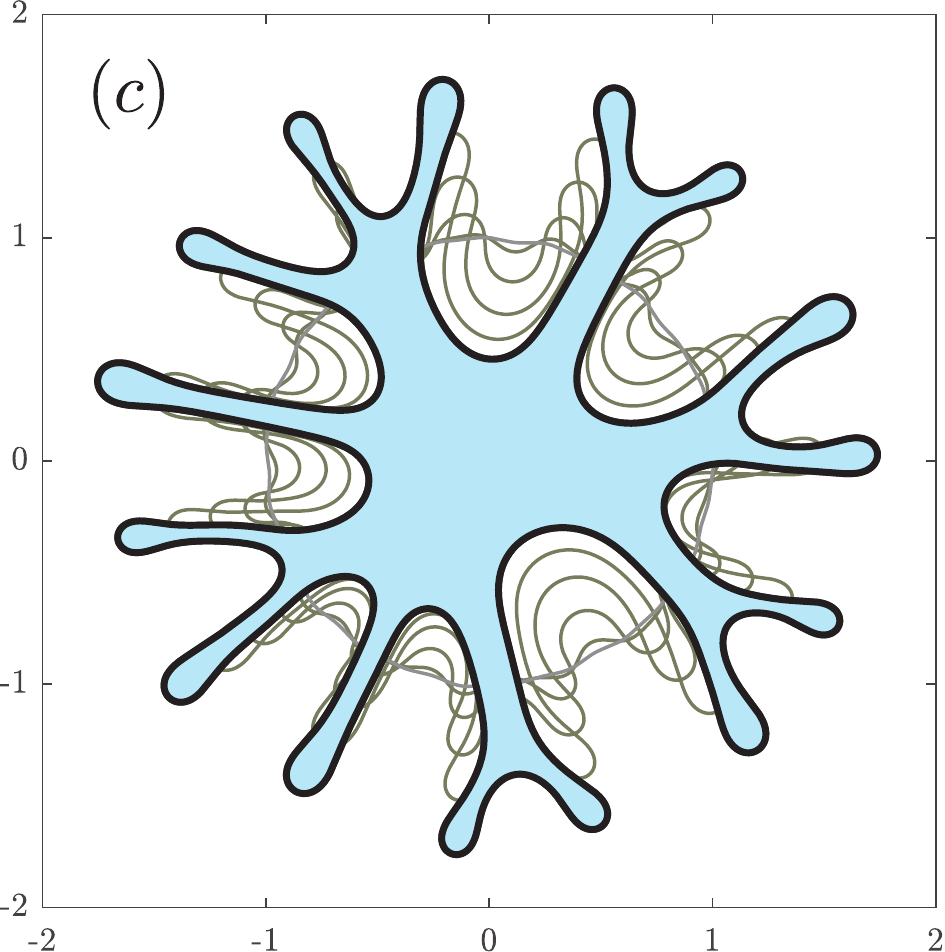} \hspace{1em}
	\includegraphics[width=0.4\linewidth]{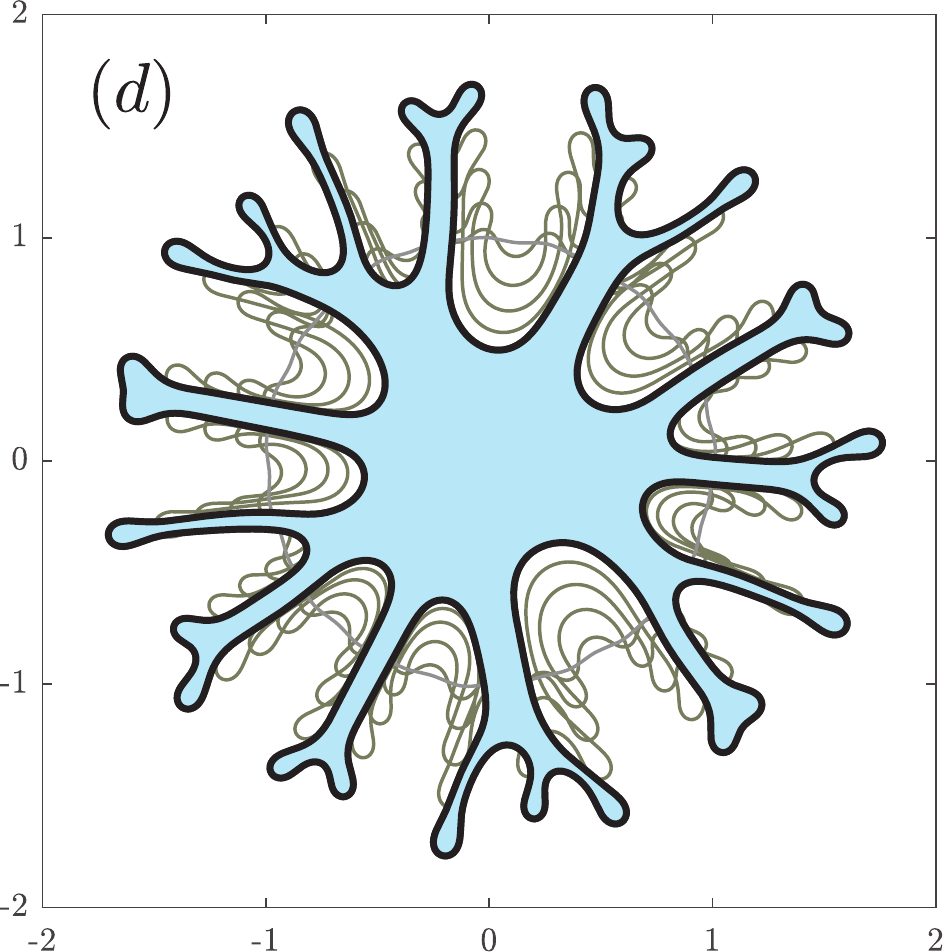}
	\caption{Numerical simulation of a rotating Hele--Shaw cell with $\omega = 1$, $Q = 0$, and $\sigma$ $(a)$ $ 10^{-2}$, $(b)$ $5 \times 10^{-3}$, $(c)$ $2.5 \times 10^{-3}$, and $(d)$ $10^{-3}$. Corresponding final time of simulations is $t = 0.61$, $0.55$, $0.425$, and $0.36$.  Initial condition for each simulation is \eqref{eq:InitialCondition2}.  Simulations are performed on the domain $-2 \le x \le 2$ and $-2 \le y \le 2$ using $500 \times 500$ equally spaced nodes.}
	\label{fig:RotatingPlates}
\end{figure}

The incorporation of the centrifugal term is straightforward. While body forces appear in the governing equation for the velocity of the viscous fluid \eqref{eq:Governing1}, scaling pressure means that the angular velocity term can be moved to the dynamic boundary condition \eqref{eq:HeleShaw2} (this can be done for any conservative body force $\vec{f}$ satisfying $\grad \times \vec{f} = \vec{0}$). The rotating Hele--Shaw cell has previously been studied numerically using boundary integral \citep{Schwartz1989} and diffusive interface \citep{Chen2014,Paiva2019} techniques. We perform simulations where the volume of viscous fluid between the plates is constant ($Q = 0$) and the Hele--Shaw plates are rotated with $\omega =1$ for different values of $\sigma$, shown in \Cref{fig:RotatingPlates}. Note that this choice of $\omega$ effectively fixes the appropriate time-scale $T$ in (\ref{eq:scaledvariables}).

For each value of $\sigma$ considered in \Cref{fig:RotatingPlates}, we find that the interface is unstable, and the fingers that develop are distinct from traditional Saffman--Taylor fingers. In particular, the fingers that develop do not appear to tip-split but instead remain constant. Additionally, the number of fingers that develop increases as the surface tension parameter is decreased such that when $\sigma = 10^{-2}$ (\Cref{fig:RotatingPlates}$(a)$), 7 fingers form, while for $\sigma = 10^{-3}$ (\Cref{fig:RotatingPlates}$(d)$), the number of fingers is approximately 21. These results are consistent with experimental results \citep{Carrillo1996} and numerical simulations \citep{Alvarez2008,Paiva2019}.

\subsection{Channel geometry} \label{sec:Channel}

In subsections~\ref{sec:Standard}-\ref{sec:Complementary}, we considered Hele--Shaw flow in radial geometry, where the bubble-fluid interface is completely immersed by an infinite amount of viscous or inviscid fluid. In this subsection, we focus on another well studied version of the Hele--Shaw cell is in channel (or rectangular) geometry, where the shape of the cell is a narrow rectangle of infinite length and width $L$. As discussed in subsection~\ref{sec:complex}, for the zero-surface-tension case, exact solutions are known to exist, which may involve a type of blow-up in finite time with a cusp forming on the boundary (as in Figure~\ref{fig:ZST}(d)).

This channel problem dates back to the work of \citet{Saffman1958}, who showed that when an inviscid bubble is injected into the channel filled with a viscous fluid, typically a single finger develops that propagates through the channel (see the numerical solution in Figure~\ref{fig:ZST}(e)). Since the work of Saffman and Taylor, extensive research has been carried out determining how the parameters of the model influence the width (relative to $L$) and speed of this finger. In particular, for a fixed injection rate, as the surface tension parameter is increased, it is established that the width and speed of the finger increase and decrease, respectively. We refer to Refs~\citet{Homsy1987,Saffman1986} (and references therein) for a comprehensive overview of the problem. In this subsection (and subsection~\ref{sec:complex}), we restrict ourselves to the classic configuration where $b$ is constant, but we note that similar to the radial problem, our scheme can easily be used to study non-standard cases, such as those in the papers \citep{Al2012,Franco2016,Thompson2014,Zhao1992}.

As with the blob problem discussed in \Cref{sec:Complementary}, it is more convenient to consider this problem in Cartesian coordinates such that $p \to p(x, y, t)$ and the interface is given by $x = f(y, t)$. Similar to the radial case (see (\ref{eq:HeleShawstandard4}), for example), the velocity of the fluid is driven by the sink term in the far-field
\begin{align} \label{eq:far-field2}
b^3 \frac{\partial p}{\partial x} \sim \frac{Q}{L} \qquad \textnormal{as} \quad x \to \pm \infty.
\end{align}
Equation \eqref{eq:far-field2} is incorporated into our finite difference stencil using a Dirichlet-to-Neumann map. We do not provide full details, but note the procedure is similar to that described in \Cref{sec:Farfield}, where we impose an artificial boundary at $x = X$, and seek a solution to \eqref{eq:HeleShaw1} of the form
\begin{align}
\hat{p}(x, y, t) = \frac{Q}{L}x + \sum_{n=0}^{\infty} A_n \textrm{e}^{-\lambda_n y} \cos \lambda_n x,
\end{align}
where $\lambda_n = 2 \pi n / L$ and $A_n$ is to be determined. Additionally, we also impose $\partial p / \partial y = 0$ on $y = \pm L/2$.

\begin{figure}
	\centering
	\includegraphics[width=0.7\linewidth]{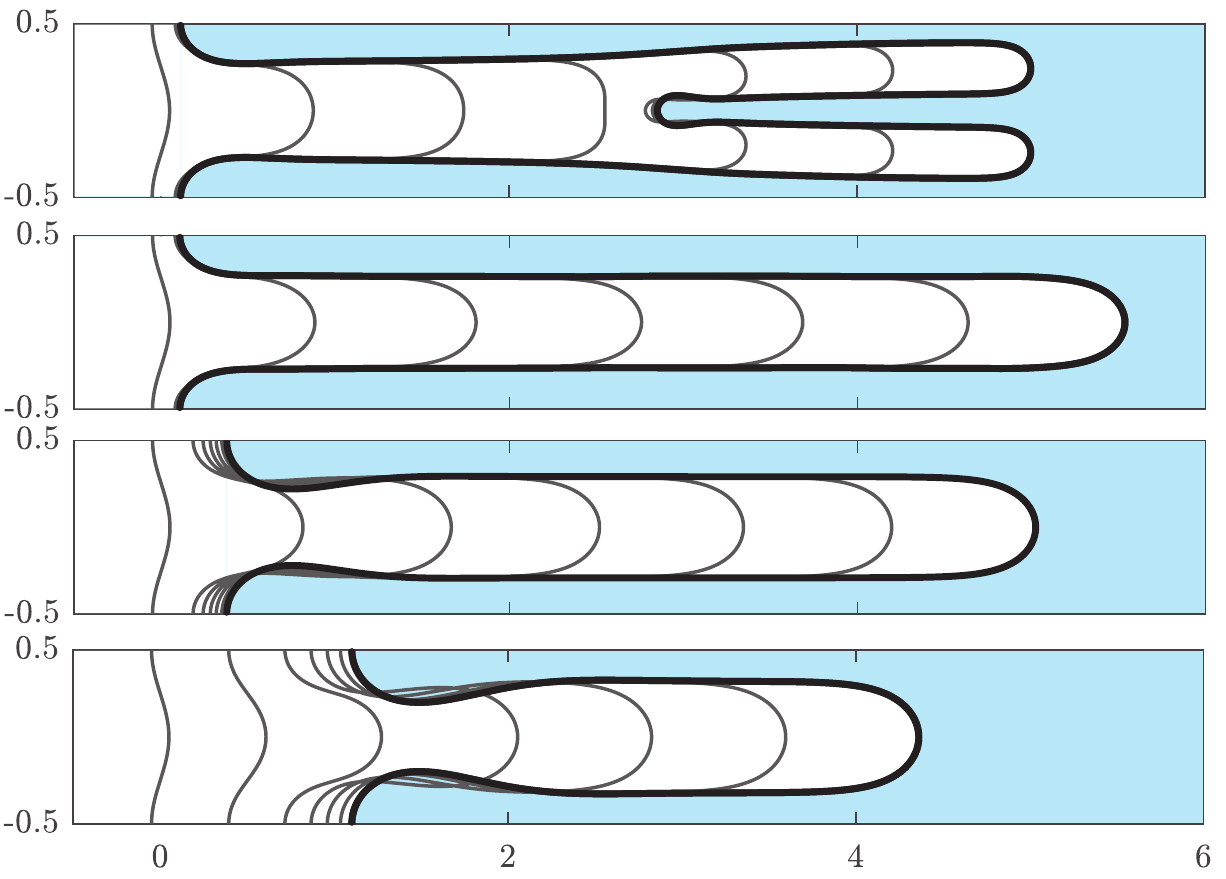}	
	\caption{Numerical simulations in channel geometry with values of the surface tension parameter (top to bottom) $\sigma = 2 \times 10^{-4}$, $5 \times 10^{-4}$, $7 \times 10^{-3}$, and $2 \times 10^{-2}$. Initial condition for all simulations is $f(x,0) = 0.05 + \cos(2 \pi y)$. Simulations are performed on the domain $-0.5 \le x \le 0.5$ and $-0.5 \le y \le 6$ using $100 \times 650$ equally spaced nodes. Solutions are plotted in time intervals of $t = 0.5$ up to $t_f = 3$.}
	\label{fig:Channel}
\end{figure}

We perform a series of simulations using the same parameters as those of \citet{Degregoria1986}, who studied this problem using a boundary-integral approach, to demonstrate our solutions are consistent with the expected behaviour. We choose the initial condition
\begin{align}
f(y, 0) = \varepsilon \cos 2 \pi y,
\end{align}
where $\varepsilon = 0.05$, and perform simulations over a range of values of $\sigma$, shown in \Cref{fig:Channel}. In this figure $L=1$, which sets the length scale $r_0$ in (\ref{eq:scaledvariables}), while $Q=1$ fixes $T$ in these equations.  For low $\sigma$ (row one of \Cref{fig:Channel}), we find that as the bubble expands, a finger grows, which is unstable and split into two. This is consistent with the results of \citet{Degregoria1986}, and this behaviour is also observed experimentally by \citet{Tabeling1987} when the injection rate is sufficiently large. For larger values of $\sigma$ (rows two to four of \Cref{fig:Channel}), a single stable finger propagates through the channel whose speed and width decreases and increases as $\sigma$ increases. Again, this behaviour is consistent with previous  experimental and numerical results \citep{Degregoria1986,Tabeling1987}.

\section{Conclusion} \label{sec:Conclusion}

In this article, we have reviewed a suite of Hele-Shaw configurations with two immiscible fluids separated by a sharp interface.  Our focus is on one-phase models, which arise by assuming one fluid is much less viscous than the other (indeed we assume the less viscous fluid is inviscid and ignore its contribution).  For the standard Hele-Shaw configuration with parallel stationary plates, we have summarised how complex variable and conformal mapping techniques can be applied to the zero-surface-tension model to deduce a variety of exact analytical results.  The three geometries we have focussed on involve a bubble expanding into a body of viscous fluid, a blob of fluid withdrawn from a point or viscous fluid displaced by an inviscid fluid in a Hele-Shaw channel.  Despite the drawbacks of Hele-Shaw models without surface tension in terms of physical applicability, these complex variable approaches are very well studied by applied mathematicians and have motivated numerous papers on moving boundary problems in general.

We have also reviewed a series of alterations to the standard one-phase Hele-Shaw model.  For these alterations applied in various combinations with the three geometries, we have presented a flexible numerical scheme based on the level set method.  We have shown that our scheme is capable of reproducing the complicated interfacial patterns that form in Hele--Shaw flow while using a uniform computational grid.  By making straightforward, appropriate adjustments to the scheme, we have been able to solve for a wide range of configurations. We have presented a selection of some of the more well-studied configurations, including the expanding bubble problem, linearly tapered plates, the withdrawal of fluid from a viscous blob, time-dependent plate gap, rotating Hele-Shaw cell, and flow in a channel geometry. For all of these configurations, we have demonstrated that our simulations compare well with previous experimental and numerical results.

While we have considered a range of different Hele--Shaw configurations in this article, this is by no means an exhaustive list. Using our numerical scheme, opportunities exist to study configurations that have not previously been considered either experimentally and numerically. For example, while the linearly tapered configuration, discussed in \Cref{sec:TaperedPlatesRadial}, has received significant attention \citep{Al2013,Anjos2018,Bongrand2018,Jackson2017}, including our own study in \citet{Morrow2019}, an open question is to determine the effect of tapering the plates for the corresponding blob problem. Our scheme could be used to gain insight the effect of the taper angle on viscous finger development when the fluid is withdrawn compared to the parallel plate case discussed in \Cref{sec:Complementary}. Further, additional physical effects on the interface between fluids could be easily included, such as kinetic undercooling \citep{Anjos2015,Dallaston2013} or dynamic wetting effects \citep{Bretherton1961,Park1984}.  Further adjustments could be made to apply the scheme to study controlling instabilities in Hele-Shaw cells with an elastic membrane \citep{Pihler2012,Pihler2013,Pihler2018} or with an external electrical field \citep{Mirzadeh2017,Gao2019b} and much more.

\subsection*{Acknowledgements}

SWM acknowledges the support of the Australian Research Council via the Discovery Projects DP140100933.  He would like to thank the Isaac Newton Institute for Mathematical Sciences, Cambridge, for support and hospitality during the programme Complex Analysis: Techniques, Applications and Computations where part of the work on this paper was undertaken.  This programme was supported by the EPSRC grant EP/R014604/1.  He is grateful for the generous support of the Simons Foundation who provided further financial support for his visit to the Isaac Newton Institute via a Simons Foundation Fellowship.

\bibliographystyle{plainnat}
\bibliography{MyBibFile}

\end{document}